\documentclass[floatfix, superscriptaddress,twocolumn, showpacs, aps, pra, 10pt]{revtex4-1}
\usepackage{amsmath}
\usepackage{natbib}
\usepackage{hyperref}
\usepackage{graphicx}
\usepackage{bbm}
\usepackage{amssymb}
\usepackage[english]{babel}
\usepackage{lmodern}
\usepackage[T1]{fontenc}
\usepackage [latin1]{inputenc}
\usepackage{isomath}
\usepackage[normalem]{ulem}
\usepackage{dcolumn}
\usepackage{color}

\graphicspath{{Figs/}}

\DeclareMathOperator{\e}{\displaystyle e}
\DeclareMathOperator{\de}{\displaystyle d}

\usepackage{amssymb}

\renewcommand{\vec}{\vectorsym}
\newcommand{\ten}{\tensorsym}
\newcommand{\mv}[1]{\langle #1 \rangle}
\newcommand{\sgn}{\operatorname{sgn}}

\begin{document}

\title{Harmonic dual dressing of  spin one-half systems}


\author{Giuseppe Bevilacqua}
\email{giuseppe.bevilacqua@unisi.it}
\affiliation{Dept. of Information Engineering and Mathematics - DIISM, University of Siena -- Via Roma 56, 53100 Siena, Italy}
\author{Valerio Biancalana}
\affiliation{Dept. of Information Engineering and Mathematics - DIISM, University of Siena -- Via Roma 56, 53100 Siena, Italy}
\author{T. Zanon-Willette}
\affiliation{Sorbonne Université, Observatoire de Paris, Université PSL, CNRS,
LERMA, F-75005, Paris, France}
\affiliation{MajuLab, CNRS-UCA-SU-NUS-NTU International Joint Research Unit,
Singapore}
\affiliation{Centre for Quantum Technologies, National University of Singapore,
117543 Singapore, Singapore.}
\author{Ennio Arimondo}
\affiliation{Dipartimento di Fisica E. Fermi, Universit\`a di Pisa -- Largo B. Pontecorvo 3, 56127 Pisa, Italy}
\affiliation{INO-CNR, Via G. Moruzzi 1, 56124 Pisa, Italy}

\date{\today}


\begin{abstract}
  Controlled  modifications of  the magnetic  response of  a two-level
  system are produced in dressed systems by one high frequency, strong
  and  non-resonant electromagnetic  field.  This  quantum control  is
  greatly  enhanced  and enriched  by  a  harmonic, commensurable  and
  orthogonally  oriented  dual  dressing,   as  discussed  here.   The
  secondary field  enables a fine  tuning of the qubit  response, with
  control parameters amplitude,  harmonic content, spatial orientation
  and phase  relation.  Our  analysis mainly  based on  a perturbative
  approach  with  respect  to  the  driving  strength,  includes  also
  non-perturbative numerical  solutions.  The Zeeman  response becomes
  anisotropic  in  a  triaxial   geometry  and  includes  a  nonlinear
  quadratic contribution.   The long-time dynamics is  described by an
  anisotropic effective magnetic field representing the handle for the
  system full engineering.  Through  the low-order harmonic mixing the
  bichromatic driving generates a synthetic static field modifying the
  system dynamics.  The spin temporal evolution includes a micromotion
  at  harmonics  of the  driving  frequency  whose  role in  the  spin
  detection is examined.  Our  dressing increases the two-level energy
  splitting, improving  the spin  detection sensitivity.  On  the weak
  field  direction  it  compensates   the  static  fields  applied  in
  different geometries.  A resonant  spin exchange between two species
  having very different  magnetic response as electron  and nucleus is
  allowed  by  the  dressing.   The   results  presented  here  lay  a
  foundation for  additional applications  to be harnessed  in quantum
  simulations.
\end{abstract}

\date{\today}

\maketitle
\section{Introduction}
Since the  early days  of quantum  mechanics, spectroscopy  probes the
energy levels of quantum systems irradiated by electromagnetic fields.
Single, bichromatic, multiple excitations are applied.  More recently,
those excitations target  the modification and the  control of quantum
properties for  the targeted object.  Quantum  variables, energies and
observables are manipulated. Within the quantum simulation effort, the
electromagnetic driving  of simple quantum systems  (mainly natural or
artificial  atoms)   generates  Hamiltonians   exhibiting  interesting
properties  hard  to  engineer  directly.  Driving  a  quantum  system
periodically in time  can profoundly alter its  long-time dynamics and
constitutes   a   versatile   scenario  to   reach   unusual   quantum
properties~\cite{GoldmanDalibard_14,BukovPolkovnikov_15,Eckardt_17}. The
dynamics associated with time-dependent  Hamiltonians is well captured
by   time-independent  effective   Hamiltonians  accounting   for  the
essential characteristics of the
modulated system.\\
\indent Within  this area  of Floquet engineering,  a pioneer  role is
played  by  the  "dressed  atom"  introduced  by  Cohen-Tannoudji  and
Haroche~\cite{CohenTannoudjiHaroche_66}, i.e., the strong driving of a
two-level  quantum  system.  The  application  of  a non-resonant  and
linearly polarized electromagnetic  field, typically radiofrequency or
microwave, allows to modify  the magnetic response.  The modifications
of the  Land\'e g-factor were  explored and applied in  various atomic
vapour
experiments~\cite{HarocheCohen_70_1,HarocheCohen_70_2,LandreCohen_70,Muskat1987,GolubLamoreaux_94,ItoYabuzaki_03,EslerTorgeson_07,Chu_11,Bevilacqua_PRA_12}. The
magnetic dressing  was studied  also with  cold atoms  and condensates
\cite{gerbier_pra_06,    Hofferberth_pra_07,BeaufilsGorceix_08}.    It
offers      a      powerful      tool     in      quantum      control
experiments~\cite{previshko_prb_15},          high          resolution
magnetometry~\cite{SwankFilippone_18},  and  spin-exchange  relaxation
experiments~\cite{Hao_pra_19}. A  key feature of this  single dressing
is  the eigenenergy  dependence on  the $J_0$  (zero-order first  kind
Bessel function), allowing a freezing of the quantum observables.  The
close connection of this $J_0$ freezing with the tunneling suppression
was pointed  out in~\cite{GrifoniHaenngi_98},  and with  the dynamical
localization in
optical lattice reviewed in~\cite{Eckardt_17}.\\
\indent  A dual  dressed  qubit  is examined  here,  more precisely  a
two-level  quantum  system  dressed  by  strong  bichromatic  harmonic
drivings based on linearly polarized dressing fields applied
along two orthogonal  axes. Its theory and  applications to different experimental configurations are presented. We compare our analyses to the previous single dressing experiments and to the dual-dressing Cs atom experiment of ref.~\cite{Bevilacqua_PRL_20} based on a weak secondary dressing. The dual dressing  quantum control is based on the combination of key ingredients  as double irradiation by the harmonic bichromatic field, large interferences in the harmonic excitation processes, strong dressed-atom driving, the generation of effective and synthetic fields  controlling the spin dynamics. Even if each of these ingredients was examined previously, their combined action was not.\\
\indent Double  irradiations, i.e., rotary saturation,  spin tickling,
and so  on, are  powerful tools in  magnetic resonance  to disentangle
complex  spectra.  In~\cite{LeskesVega_10}  a  low order  perturbation
analysis of the Floquet treatment examines those regimes.  Experiments
in quantum optics have studied  the bichromatic driving as reviewed in
\cite{FicekFreedhoff_00,SilveriParaoanu_17}.    The    attention   was
focused  mainly  on  the  absorption  of  atomic/molecules,  and  more
recently          also           of          artificial          atoms
in~\cite{PeirisMuller_14,SataninNori_14,ForsterKohler_15,PanHan_17,DaiYu_17}. Multifrequency
excitation has  received a  large theoretical  attention based  on the
continued fraction of
matrices and complex numerical solutions~\cite{HoChuTietz_83,Chu_85,PoertnerMartin_20}.     Incommensurable dual driving is  used by~\cite{GarridoPerrinLorent06} for the evaporative  cooling control, by~\cite{TrypogeorgosSpielman_18,AharonRetzker_19} for the shield of optical clock transitions from magnetic static field, and for the probing of atom-photon interactions in~\cite{LukschFoot19}. \\
\indent  The interference  in the  driving  of a  two-level system  by
several  harmonic  fields  was  examined  in  the  magnetic  resonance
experiments of  refs.~\cite{YabuzakiOgawa_74,TsukadaTomishima_81}. The
dual modulation driving in an optical lattice
clock by ref.~\cite{LuChang_21}  evidences both  the interference and the driving phase  role.\\
\indent The
bichromatic harmonic excitation is important  for  the physics of ultracold atoms in optical lattices. The generated tunnelling suppression  in a lattice dual well, i.e., a generalization of the $J_0$ freezing,  is examined  theoretically in~\cite{FarrellyMilligan_93,KarczmarekIvanov_99} and experimentally in~\cite{GoergEsslinger_19}. That driving allows also to engineer the nearest-neighbor interactions in~\cite{ZhaoKnolleMintert_19} and  the dissipation processes in \cite{ViebahnEsslinger2020}.\\
\indent  The dressed  atom  modification by  a second  electromagnetic
field with its frequency quasi resonant with the dressing one, was
briefly explored  by ref.~\cite{Loginov_83} in 1983. \\
\indent An effective Hamiltonian is often used to analyse the dynamics
of                            driven                           quantum
systems~\cite{GoldmanDalibard_14,BukovPolkovnikov_15,Eckardt_17}.
Ref~\cite{KarczmarekIvanov_99}   derives   such  Hamiltonian   for   a
tight-binding model of the bichromatic driving tunnelling suppression.
For  the atomic  trapping  by rf  dressed  adiabatic potentials,  that
Hamiltonian describes  the synthetic  fields created  by commensurable
bichromatic  or multiple  drivings~\cite{HarteFoot_18}.  An  effective
Hamiltonian for  the action  of the bichromatic  drive on  a two-level
system is  introduced by~\cite{SaikoFedaruk_18}.  As an  equivalent of
our  synthetic  field, the  bichromatic  harmonic  driving produces  a
rectified transport when  applied to external degrees  of freedoms, as
for                          quantum                          ratchets
in~\cite{GoychukHaenngi_98,LebedevRenzoni_09,HaenggiMarchesoni_09}.
That effective Hamiltoninan  is examined by~\cite{WitthautWeitz_11} in
an experimental driving of ultracold atoms. The rectified transport is
studied in~\cite{KohlerStauber_20}  for a graphene model  based on the
coupling between driven spin and electron
momentum. \\
\indent Our analysis show that a dressed spin experiences a micromotion, i.e., a temporal evolution at harmonics of the driving frequency, as for other periodically driven quantum systems in~\cite{GoldmanDalibard_14}. From the  experimental point of  view, the micromotion has received a large attention for trapped ions, and for atoms in optical lattices by~\cite{DesbuquoisEsslinger_17,ArnalGueryOdelin_20}. We derive that in single and dual spin dressing  the micromotion is composed by two separate components, one given by a gauge transformation and the second one by a kick operator. Their role on the dressed evolution and detection is discussed.\\
\indent Owing to  the combination of the  above ingredients, important
original  features   are  associated   to  this   configuration.   The
eigenvectors and eigenvalues  of the dressed spin are  described by an
effective magnetic Hamiltonian, providing  a simple description of the
spin dynamics.  Its terms are finely tuned by the dressing amplitudes,
their relative phase, the spatial  orientation of the dressing fields,
and the  driving harmonic  order.  At weak  static fields,  the lowest
order  Zeeman  coupling with  the  static  field  is described  by  an
effective tensorial Land\'e g-factor, with a triaxial magnetic control
analog  to   the  magnetic   anisotropies  appearing   in  solid-state
materials.  A weak tensorial  nonlinear Zeeman coupling contributes to
the effective  Hamiltonian. The  dual dressing creates  an arbitrarily
oriented synthetic static  magnetic field, i.e., even in  absence of a
real  external field.   The  underlying process,  a nonlinear  optical
rectification of  the dressing fields, is  equivalent to light-shifts,
even higher  order ones, due to  the combined action of  those fields.
The   effective  Hamiltonian   components   are   determined  by   the
interferences  between absorption/emission  processes of  both fields,
the interferences enhanced by the low-harmonic order of
the  harmonic  driving. These features introduce additional original degrees of freedom in the quantum control of the spin dynamics. \\
\indent Several  applications for quantum technologies  are presented.
For instance our dressing geometry generates effective magnetic fields
one thousand times larger than the externally applied static field, or
produces   an   arbitrary   compensation   of   external   arbitrarily
orientedstatic  fields, with  a  reduced sensitivity  to the  dressing
parameters.  Even if  our analysis concentrats on  a two-level system,
the presence  of external levels  does not modify the  listed features
because the dressing is based  on a non-resonant excitation. All these
tools constitute an exceptional handle in a wide
range of directions, from selective spectroscopic detection to quantum simulation and computation.\\
\indent This  paper is  structured as  follows. After  introducing the
Hamiltonian and the associated  Floquet engineering, Sec.  II presents
the  dual   dressing  main  features.   Sec.    III  investigates  the
perturbation  regime with  one  dressing field  larger  than both  the
static field and  the secondary dressing one.  Sec.   IV discusses the
connection  between   experimental  detection  and   system  dynamics,
including  the micromotion.   Sec.  V  explores original  applications
allowed  by   the  dual-dressing.   A  final   Section  concludes  our
work. Short Appendices
report mathematical derivations. \\
\section{Hamiltonian and effective fields}
 \subsection{The Hamiltonian}
\begin{figure}[!!b]
   \centering
    \includegraphics [angle=0, width= 0.6\columnwidth] {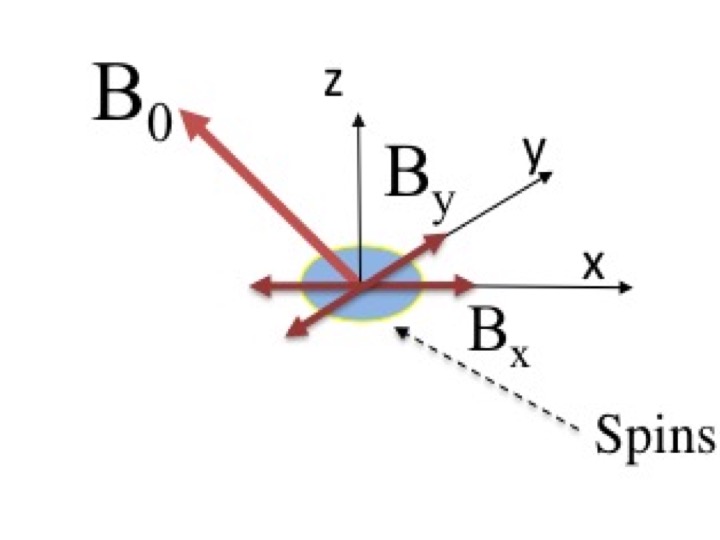}
    \caption{Schematic  of an atomic spin  system dressed
      by the $B_x$ and $B_y$ oscillating fields, in presence of static
      field  $\vectorsym{B}_0$  arbitrarily  oriented. Initially the  spins  are
      optically pumped into an eigenstate.}
 \label{fig:setup}
\end{figure}

 \indent  We consider  a spin  $1/2$ system  (either real  or
 artificial  atom) interacting  with static  and oscillating  magnetic
 fields, as schematized in Fig.~\ref{fig:setup}  for an atomic physics setup.  The  spin-field coupling  is determined by  the gyromagnetic
 ratio $\gamma=g\mu_B/\hbar $  with $g$ the  Land\'e  factor and $\mu_B$  the Bohr
 magneton. The $\vectorsym{B}_0$ static magnetic field  has components
 $B_{0j}$  on  the  $j  =  (x,  y,  z)$  axes.  The spin  is
 driven  by  two  time-dependent  and  periodic  fields   oriented  along the  $x$  and  $y$ axes,  respectively,  with $B_i$  the maximum value and $s_i(t)$
the  time  dependence,  where $(i=x,y)$.  The
 Hamiltonian is
\begin{equation}
  \label{eq:hami:ini}
  H =\frac{\gamma}{2} \left[ \vectorsym{B}_0\cdot \vectorsym{\sigma} +
 B_x \, s_x(t) \,\sigma_x + B_y \, s_y(t) \,\sigma_y\right],
\end{equation}
with $\sigma$ the Pauli matrices. We concentrate  on  the
following time dependencies:
\begin{equation}
s_x(t)=\cos(\omega t), \,\quad  s_y(t)=\cos(p\omega t+\Phi),
\label{eq:sinusoidal}
\end{equation}
with $p$  an integer and $\Phi$  the phase difference of  the harmonic
fields. The role of the $s_x$ initial phase is briefly analyzed within the detection Section.\\
\indent  The  angular frequency $\omega$ is taken as the frequency unit.. We introduce the  dimensionless time $\tau=\omega t$ and the dimensionless $\vectorsym{\omega}_0     =     \gamma     \vectorsym{B}_0/\omega$ and 
$\Omega_i = \gamma B_i/\omega$, the bare Larmor frequency and the Rabi frequencies, respectively. The
$U(\tau)$ time evolution operator of the Hamiltonian results
\begin{equation}
  \label{eq:def:U}
  i \dot{U}(\tau) = \frac{1}{2}\left[ 
   \vectorsym{\omega}_0\cdot \vectorsym{\sigma}
  + \Omega_x\, s_x(\tau) \,\sigma_x
  + \Omega_y \, s_y(\tau) \,\sigma_y
  \right] U(\tau),
\end{equation}
and within this rewriting the spin-field coupling is described by an effective g-Land\'e factor equal one. \\
\indent The                                                            Floquet
theorem~\cite{GoldmanDalibard_14,BukovPolkovnikov_15,Eckardt_17}
allows us to write 
\begin{equation}
  \label{eq:U:floqu}
  U(\tau) = {\cal M}(\tau) \e^{-i \Lambda \,\tau} 
\end{equation}
with   ${\cal  M}(0)=1$  and ${\cal  M}(\tau+2\pi)  =  {\cal
  M}(\tau)$.  The  ${\cal M}$ operator describes  the spin micromotion, and
$e^{-i   \Lambda\tau}$    represents   the    stroboscopic   evolution
operator.  For  single dressing the  ${\cal M}$ operator was derived in~\cite{BaroneNarcowich_77} up to the
fourth order perturbation in the dressing amplitude.\\
 \begin{figure}[!!b]
   \centering
    \includegraphics [angle=0, width= 0.9\columnwidth] {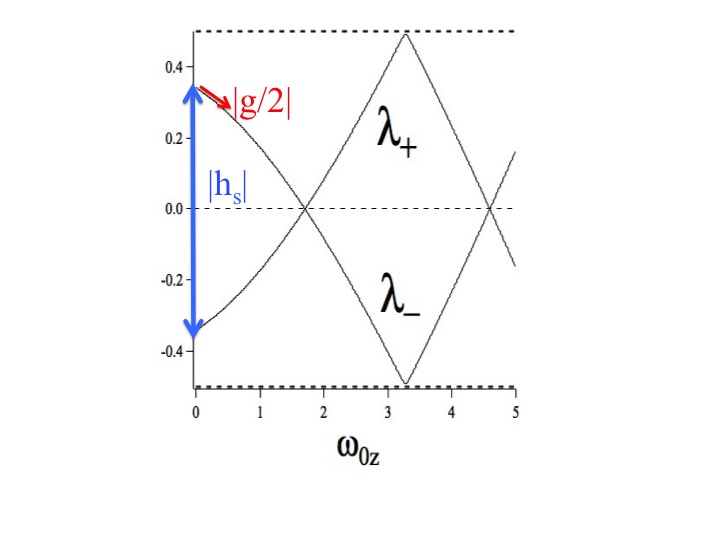}
    \caption{First   Brillouin  zone   $\lambda_+,\lambda_-$
      eigenvalues   vs  $\omega_{0z}$.   The
      $|\vec{h}_s|$ modulus  is  the energy  separation at
      zero field.  The $|\vec{g}|$ amplitude  is the absolute value for the derivative  of that
      separation. Parameters     $\omega_{0x}=\omega_{0y}=0$,
      $\Omega_{x}=5.11,      \Omega_{y}=3$,      $p=1$,      $\Phi=\pi/2$.   For this case, at $\omega_0\simeq 0$,  $\Omega_L=\lambda_+-\lambda_-$ is larger than $\omega_0$.  }
\label{fig:eigenvalues}
\end{figure}
\subsection{Effective field}
\indent  The  spin  dynamics  is  determined  by  the  $\Lambda$  time
independent  Floquet Hamiltonian.  This  matrix is  not unique  since, for a  given $U$  operator, one  can
subtract  multiples  of  the  $\omega$  frequency  from  its  diagonal
elements and compensate  by multiplying ${\cal M}(\tau)$  with a diagonal
matrix.  The  $\Lambda$
matrix is written as
\begin{equation}
  \Lambda = \frac{1}{2} \, \vectorsym{h} \cdot \vectorsym{\sigma},
  \label{eq:effectivefield}
  \end{equation}
with its  eigenvalues limited  to the  $[-1/2,  1/2]$ first  Brillouin  zone, as in Fig.~\ref{fig:eigenvalues}.  The $\vectorsym{h}$ vector, measured  in energy units, represents an
  effective  magnetic field.  The $\lambda_{\pm}$ eigenvalues result
  \begin{equation}
  \label{eq:floqu:expo}
 \lambda_{\pm} =
  \pm \frac{ \sqrt{\vectorsym{h} \cdot \vectorsym{h} }}{2}
\end{equation}
\indent The effective magnetic field produces an energy splitting
 described by the dressed $\Omega_L$ Larmor
frequency given by
\begin{equation}
  \label{eq:defWL}
  \Omega_L =\lambda_+ - \lambda_- =  \sqrt{\vectorsym{h} \cdot \vectorsym{h} }
\end{equation}
As in
Fig.~\ref{fig:eigenvalues},  the  Larmor 
precession  frequency 
may reach the maximum value of one, or  in natural  units the
$\omega$   value.    Therefore,  at low  $\omega_0$ values $\Omega_L$ 
may become larger    than     the  bare Larmor frequency, as in the figure.\\
\indent    The   eigenstates    of   the    $\Lambda$   operator    of
Eq.~\eqref{eq:effectivefield} correspond to the spin oriented parallel
or antiparallel to $\vectorsym{h}$ vector, i.e., the orientation
unit vector  $\vectorsym{u}=\vectorsym{h}/\Omega_L$.   The spin  precession
takes  place in  the  plane orthogonal  to  $\vectorsym{u}$.  The 
effective field definition may be rewritten as
 \begin{equation}
  \Lambda = \frac{1}{2} \, \Omega_L \vectorsym{u} \cdot \vectorsym{\sigma}.
  \label{eq:effectivefield2}
  \end{equation}\\ 
\indent   The     effective  field  $\vectorsym{h}$  depends  on   both
$\vectorsym{\omega}_0$ and  $\Omega_i$, with $(i=x,y)$.  In  the small
$\vectorsym{\omega}_0$  limit  of  our   interest,  we  introduce  the
following Taylor expansion:
\begin{equation}
  \label{eq:taylor:3ord}
   \begin{split}
  &h_i(\vec{\omega}_0, \Omega_x,\Omega_y) = h_{s,i} \\
  &+ \sum_{j=x,y,z}\ten{g}_{ij} \;\omega_{0j} 
  + \frac{1}{2} \sum_{j,k=x,y,z}\ten{f}_{ijk} \;\omega_{0j}\;\omega_{0k} + \ldots
 \end{split}
 \end{equation}
 The $(\Omega_x,\Omega_y) $ parameter dependence applies to
 all the right side quantities. A similar
 expansion of  the Floquet eigenvalue  at the  zero magnetic
 field was derived for the
 Bloch-Siegert shift in~\cite{HannafordPeggSeries_73}. \\
 \indent  The  zeroth-order  vector  $\vectorsym{h}_s$  synthetic  (or
 fictitious)  field  represents a static field acting on the spin also in absence of the externally applied static field. It is given  by  the  $\lambda_+- \lambda_-$  energy
 separation     at    ${\vec     \omega}_{0}=0$     as    shown     in
 Fig.~\ref{fig:eigenvalues}(a).  The $\vectorsym{h_s}$ field  is equivalent
 to  the light  shifts~\cite{CohenTannoudjiGueryOdelin_94} and  to the
 generalized Bloch-Siegert  shifts~\cite{SilveriParaoanu_17}.  Here it
 contains high order shifts due to  the combined action of the harmonic driving
 fields. The synthetic  field is generated  by a nonlinear
 optical rectification process of the harmonic driving fields. \\
 \indent Within  the Taylor expansion of Eq. \eqref{eq:taylor:3ord},  the first-order Jacobian
 matrix denoted as $\ten{g}$ is given by
\begin{equation}
  \label{eq:def:G}
  \ten{g}_{ij} \equiv \left.  \frac{ \partial h_i}{\partial \omega_{0j}}
  \right|_{\vectorsym{\omega}_0=\vectorsym{0}}.
\end{equation}
\indent This tensor produces a linear Zeeman effect and represents the
generalization   of   the   diagonal   Land\'e   g-factor   introduced
by~\cite{LandreCohen_70}. Its modulus  is schematically represented in
Fig.~\ref{fig:eigenvalues}(a).  The g-factor  as a second-order tensor
appears  in  electron spin  resonance  in  solids.  There  the  tensor
principal  axes  (tensor  eigenstates)  are determined  by  the  local
fields.  Here  the $\ten{g}$ principal  axes define the  basic spatial
directions of the spin
magnetic response.  The $Pv_i$ for $i=(1,3)$, principal values (tensor
eigenvalues) contain phases modifying the spin temporal evolution.

 The  Taylor expansion  leads  to  the $\ten{f}$  second-order
gradient tensor defined by
\begin{equation}
  \label{eq:def:3ord:tens}
  \ten{f}_{ijk} \equiv  \left.  \frac{\partial^2 h_i}{\partial
      \omega_{0j}\;\partial \omega_{0k} }
  \right|_{\vectorsym{\omega}_0=\vectorsym{0}}.
\end{equation}
This  tensor  corresponds  to   a  nonlinear  and  anisotropic  Zeeman
effect. It is equivalent to the  Hessian control for the spin response
in ~\cite{ShenRabitz06} and in geophysics  to the full gradient tensor
of a magnetic field~\cite{SuiClark_17}.

\section{Perturbative and numerical solutions}
\label{sec:pert:sol}

A perturbative  approach allows us  to derive  in Subsections A  and B
general analytical  expressions for  the above defined  quantities for
generic  $s_x,s_y$ time  dependencies.  In  the following  Sections we
focus on the  cosine driving and presents plots for  the values of the
components for effective and synthetic fields, $\vec{g}$ and $\vec{f}$
tensors.  In that  context the results of numerical  solutions for the
dual-dressing Hamiltonian are also presented.

\subsection{Frequency-modulated rotating field}

The case  $\Omega_x \gg  \Omega_y$ is considered  here, always  in the
limit of a weak $\vec{\omega}_0$ static field. The opposite case would
just correspond to swapping the axis  labels.  We define the $x$ phase
accumulated  by the  spin  for a  periodic  $s_x(\tau)$ driving,  more
general than in Eq.~\eqref{eq:sinusoidal}
\begin{equation}
  \label{eq:def:phi}
  \varphi_x(\tau)  = \Omega_x  \int_0^{\tau} s_x(\tau  ') \de\ \tau'.
\end{equation}
\indent For  a rotating dressing field  the interaction representation
in  a  frame rotating  about  the  static  field axis  simplifies  the
description.  For  our case  of  weak  static  fields such  usual  RWA
approximation  is not  valid.  Therefore  we introduce  an interaction
representation with  respect to the strong  $\Omega_x$ dressing field,
the time evolution operator being factorized as
\begin{equation}
  \label{eq:U:ri}
U(\tau) = \e^{-i \varphi_x  \sigma_x/2} \, U_I(\tau).  
\end{equation}
\indent This  gauge transformation represents  a unitary change  to a
reference  frame rotating  about the  $x$ axis  with a  rotation angle
presenting  a   non-trivial  time  dependence.   For  the  $s_x$
sinusoidal  time  dependence,  the rotation  angle  experiences  a
frequency-modulated       rotation       (FMR),       as       denoted
in~\cite{PeggSeries_70}.\\ 
  \indent  The $U_I$ time evolution is given by
\begin{equation}
  \label{eq:def:UI:din}
  i \dot{U}_I(\tau) = \frac{1}{2}\e^{i \varphi_x \sigma_x/2}
  \left[
    \vec{\omega}_0\cdot\vec{\sigma} +
    \Omega_y s_y(\tau) \sigma_y
  \right]
  \e^{-i \varphi_x \sigma_x/2} \; U_I(\tau).
  \end{equation}
  \indent We  apply the  explicit expression  for the  exponentials of
  Pauli  matrices,  use   the  Baker-Cambell-Hausdorff  relation,  and
  manipulate the result with the  commutation rules of those matrices.
  After this algebra the $U_I$ operator is rewritten as
\begin{equation}
  \label{eq:def:din}
  i \dot{U}_I(\tau) = 
   \frac{1}{2}\left[ \vec{h}^{FMR}(\tau) \cdot \vec{\sigma}\right]\; U_I(\tau),
\end{equation}
with  $\vec{h}^{FMR}(\tau)$  given by
\begin{widetext}
\begin{equation}
  \label{eq:def:m}
  \vec{h}^{FMR}(\tau) = 
  \begin{pmatrix}
    \omega_{0x}\\
    \omega_{0y} \cos(\varphi_x(\tau))  + \omega_{0z} \sin(\varphi_x(\tau))
    + \Omega_y s_y(\tau) \cos(\varphi_x(\tau)) \\
    \omega_{0z} \cos(\varphi_x(\tau))  - \omega_{0y} \sin(\varphi_x(\tau))
    - \Omega_y s_y(\tau) \sin(\varphi_x(\tau)) 
  \end{pmatrix}.
\end{equation}
\end{widetext}
The vector  $\vec{h}^{FMR}(\tau)$ represents the  time-dependent field
acting  on the  spin in  the FMR  reference frame.   Because both  the
$s_x(\tau)$    signal   of    Eq.   \eqref{eq:sinusoidal}    and   the
$\varphi_x(\tau)$ are periodic functions, also the
$\vec{h}^{FMR}$ vector is periodic. 

The  time evolution of Eq.~\eqref{eq:def:din}  being periodic,
the Floquet theorem allows us to write the $U_I(\tau)$ operator as
\begin{equation}
  \label{eq:form:U:floquet}
  U_I(\tau) = \e^{-i {\cal K}(\tau)} \; \e^{-i \Lambda \; \tau}
\end{equation}
where the kick operator ${\cal K}(\tau)$  is defined as in   the modulated optical lattice descriptions~\cite{GoldmanDalibard_14,BukovPolkovnikov_15,Eckardt_17}. The FMR kick operator is periodic and satisfies ${\cal K}(0) = 0$. Because the $U(\tau)$ lab frame operator is derived from the $U_I(\tau)$ FMR frame operator using Eq.\eqref{eq:U:ri}, the $\Lambda$ matrix of Eq.~\eqref{eq:form:U:floquet} is the stroboscopic operator introduced in Eq.\eqref{eq:U:floqu}.  
\subsection{First and second order solutions}
\label{sec:1ord:sol}
\indent  The $U_I(\tau)$ evolution in the FMR frame allows us to  derive from the  time-dependent $\vec{h}^{FMR}(\tau)$ of Eq.\eqref{eq:def:m} an analytical expression of the effective static  field   $\vec{h}$  defined in
Eq.~\eqref{eq:effectivefield},  always in the limit of both $\Omega_y$  and
$\vectorsym{\omega}_0$  small  parameters.   From appropriate  time
averages       of        $\vec{h}^{FMR}(\tau)$   of      
Appendix~\ref{sec:FMReffectiveFields}, we derive  the first and second
order perturbation $\vec{h}$ expressions defined as 
\begin{equation}
  \label{eq:h:pertur}    
    \vec{h}  \approx  \vec{h}^{(1)} + \vec{h}^{(2)}.
\end{equation} 
Using Eq.~\eqref{eq:def:1e2ord} the first order, given by the $\vec{h}^{FMR}(\tau)$ FMR time average, results 
\begin{equation}
  \label{eq:def:h:mv}
  \vec{h}^{(1)} = 
  \begin{pmatrix}
    \omega_{0x}\\
    \omega_{0y} \mv{\cos\varphi_x}  + \omega_{0z} \mv{\sin\varphi_x}
    + \Omega_y \mv{ s_y \cos\varphi_x} \\
    \omega_{0z} \mv{\cos\varphi_x}  - \omega_{0y} \mv{\sin\varphi_x}
    - \Omega_y \mv{ s_y \sin\varphi_x} 
  \end{pmatrix}.
\end{equation}
\indent  Using the  $\alpha_n$ and  $\beta_n$ definitions  in Appendix
Eqs.~\eqref{eq:def:csp:coeffs}, we  obtain the following  first order
synthetic field and g-tensor: 
\begin{subequations}
  \label{eq:h_synt:1o}
  \begin{align}
    \vec{h}_s^{(1)} &= \Omega_y
   \begin{pmatrix}
     0\\
    \phantom{-}\mathrm{Re}(\beta_0) \\
   -\mathrm{Im}(\beta_0)
   \end{pmatrix}, \\
    \ten{g}^{(1)} &=
        \begin{pmatrix}
         1 &0 & 0 \\
        0 & \phantom{-}\mathrm{Re}(\alpha_0) & \mathrm{Im}(\alpha_0) \\
        0 & -\mathrm{Im}(\alpha_0) & \mathrm{Re}(\alpha_0) 
        \end{pmatrix} \nonumber \\
  & =
    \begin{pmatrix}
   1 &0 & 0 \\
   0 & |\alpha_0| & 0 \\
   0 & 0 & |\alpha_0|
 \end{pmatrix}
   \begin{pmatrix}
    1 &0 & 0 \\
    0 & \phantom{-}\cos(\eta_0) & \sin(\eta_0) \\
    0 & -\sin(\eta_0) & \cos(\eta_0) 
  \end{pmatrix},
  \end{align}                   
\end{subequations}
where $\tan \eta_0 = \mathrm{Im}(\alpha_0)/\mathrm{Re}(\alpha_0)$. 
At this order there is no contribution to $\ten{f}$.\\
\indent  This  $\vec{g}^{(1)}$  expression  represents  a  tensor 
non-diagonal  and  non-symmetric,  corresponding  to  a  contraction  and
rotation of the spin response.  Its  $Pv_{i}$ $(i=1,3)$ principal values are $\left[1,
  \mathrm{Re}(\alpha_0)\pm i \mathrm{Im}(\alpha_0)\right]$.  These complex conjugates ones correspond to tensor eigenvectors  in the
$(y,z)$ plane experiencing  different phase shift with  respect to the dressing fields.\\ 
\indent    The second  order  synthetic  field derived  
from  Eq.~\eqref{eq:mxM} contains a single component  as
\begin{equation}
  \label{eq:h20:synte}
  \vec{h}_s^{(2)} = \frac{\Omega_y^2}{2}\begin{pmatrix}
    Q_{x}\\0\\0 
  \end{pmatrix},
\end{equation}
with 
\begin{equation}
\label{eq:qx}
 Q_{x}  =
  \sum_{n\neq 0}
  \frac{|\beta_n|^2 - 2\, \mathrm{Re} ( \beta_0^* \beta_n) }{n}.
\end{equation}
The second order expression of the  g-tensor reads as
\begin{equation}
  \label{eq:g2ord}
  \ten{g}^{(2)} =
  \frac{\Omega_y}{2}
  \begin{pmatrix}
    0 & Q_{xy}& Q_{xz}\\
    Q_{yx}& 0 & 0\\
    Q_{zx} & 0 & 0 
  \end{pmatrix},
\end{equation}
where
\begin{eqnarray}
   \label{eq:def:Qij}
  Q_{xy} & = 2\, \mathrm{Re} \bigg( \sum_{n\neq 0}
             \frac{\beta_n \alpha_n^*-
             \beta_0 \alpha_{n}^* - \alpha_0^* \beta_n}{n} \bigg), \nonumber \\
    Q_{xz} &= - 2 \,\mathrm{Im} \bigg( \sum_{n\neq 0}
             \frac{\beta_n \alpha_n^*-
             \beta_0 \alpha_{n}^* - \alpha_0^* \beta_n}{n} \bigg), \nonumber\\
    Q_{yx} & =  2 \,\mathrm{Re} \bigg( \sum_{n\neq 0}
             \frac{\beta_n}{n} \bigg), \nonumber\\
  Q_{zx} &= -2 \, \mathrm{Im} \bigg( \sum_{n\neq 0} \frac{\beta_n}{n} \bigg)
             . 
\end{eqnarray}
At  this order  the  $\ten{f}^{(2)}_{i,j,k}$  tensor components,  with
$(i=x,y,z)$, are the following ones, respectively: 
\begin{equation}
  \label{eq:Y1}
 \left[  \begin{pmatrix}
    0 & 0 & 0 \\
    0 & q_0 & 0 \\
    0 & 0 & q_0  
  \end{pmatrix}, 
   \begin{pmatrix}
    0 & q_s & -q_c  \\
    q_s & 0 & 0 \\
    -q_c& 0 & 0  
   \end{pmatrix},
 \begin{pmatrix}
    0 & q_c & q_s \\
    q_c & 0 & 0 \\
    q_s& 0 & 0  
  \end{pmatrix} \right],
\end{equation}
where
\begin{eqnarray}
  \label{eq:def:qc:qs}
  q_0 &=\phantom{-}   \sum_{n\neq0}\frac{|\alpha_n|^2 - 2 \,\mathrm{Re}(\alpha_n \alpha_0^*)}{n}, \\
  q_c &= - \,\mathrm{Im}\bigg(
        \sum_{n\neq 0}\frac{\alpha_n}{n}\bigg),\\
  q_s & = \phantom{-} \,\mathrm{Re}\bigg(
        \sum_{n\neq 0}\frac{\alpha_n}{n}\bigg). 
\end{eqnarray}
At  this order  the $\ten{f}$  tensor components  depend only  on the
 strong  field  variables,  as  the  $\vec{g}$  tensor  at  the  first
 order. \\ 
 \begin{figure}
   \centering
    \includegraphics [angle=0, width= 0.95\columnwidth] {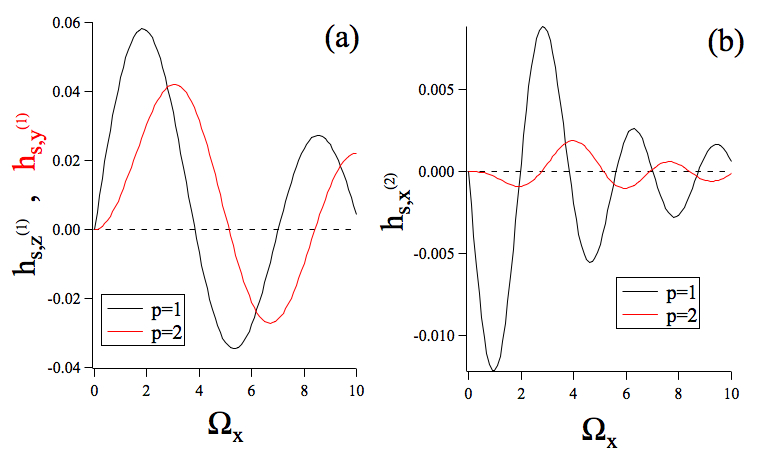}
      \includegraphics [angle=0, width= 0.9\columnwidth] {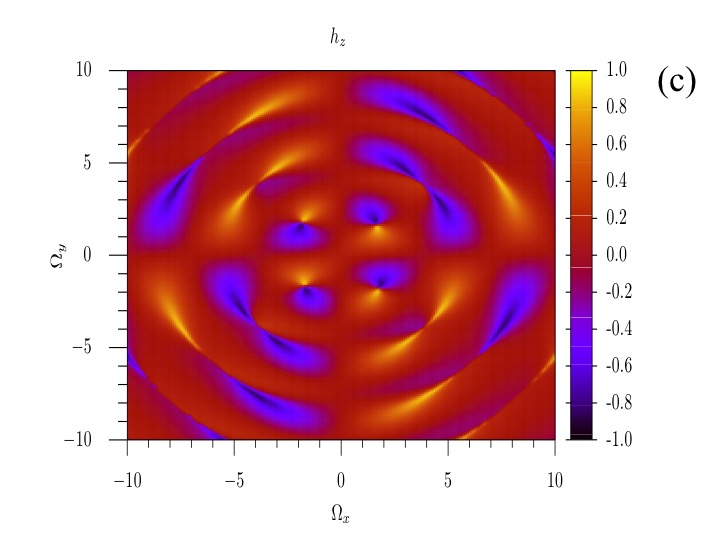}
        \caption{Cosine driving synthetic/effective    fields   for   $\omega_{0z}=0.1$    and
          $\omega_{0x}=\omega_{0y}=0$.   In  (a) and (b) synthetic fields derived from the perturbation treatment  vs  $\Omega_x$ at $\Omega_y=0.1$. In (a) black line  for  $h^{(1)}_{s,z}$ at $p=1$ and  $\Phi=\pi/2$;
          red  line  $h^{(1)}_{s,y}$   at  $p=2$  and  $\Phi=\pi/6$.  In (b) second order   $h^{(2)}_{s,x}$
          vs  $\Omega_x$   at  $\Omega_y=0.2$ and $(p=1,2)$.    In (c) numerical results for the $h_{z}$ effective field in a $(\Omega_x,\Omega_y)$ two-dimensional (2D)  plot
                  at  $p=1$ and $\Phi=\pi/2$.} 
 \label{fig:hzplots}
\end{figure}

\subsection{Cosine signals}
\subsubsection{Phase controlled cosines}
\label{sec:simple:cos}
\begin{figure}
   \centering
    \includegraphics [angle=0, width= 0.9\columnwidth] {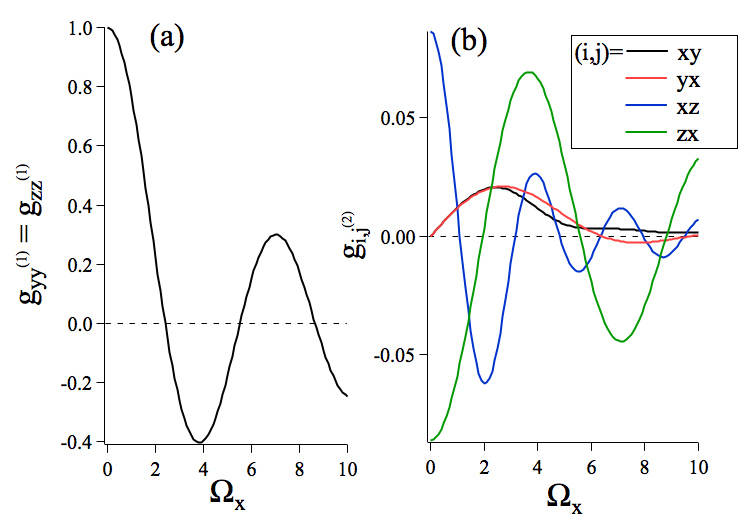}
      \includegraphics [angle=0, width= 0.6\columnwidth] {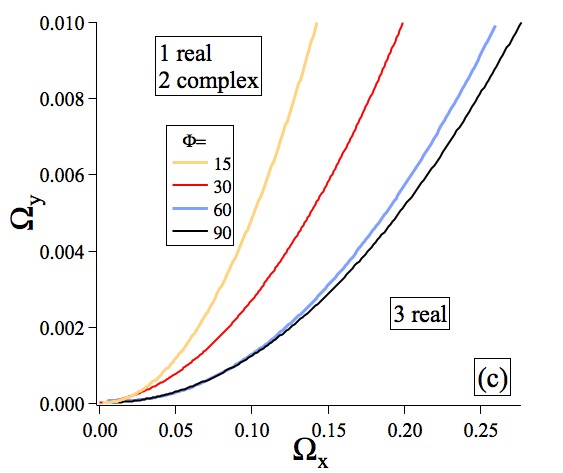}
        \caption{Cosine driving results for the  $\ten{g}$ components. In     (a)    $\ten{g}^{(1)}_{xx}=\ten{g}^{(1)}_{yy}$
          components vs $\Omega_x$, given by the $J_0$ Bessel function
          and independent  of $\Omega_y$. In  (b) $\ten{g}^{(2)}_{ij}$
          tensor  components  vs   $\Omega_x$  at  $\Omega_y=0.2$, $p=1$ and
          $\Phi=\pi/3$ .    In  (c) perturbation derived lines separating  the
          $(\Omega_x,\Omega_y)$  regions  of real principal values, below the parabole, and of real/complex principal values, above the parabole.  Parameters $p=1$
          and   different  $\Phi$   values. } 
 \label{fig:gplots}
\end{figure}

The $(x,y)$  cosine driving  case of Eq.~\eqref{eq:sinusoidal}  with a
controlled  $\Phi$  phase difference  between  the  driving fields  is
discussed here.  The FMR  accumulated phase  of Eq.~\eqref{eq:def:phi}
becomes
\begin{equation}
  \label{eq:sd:prl}
  \varphi_x(\tau) = \Omega_x\sin(\tau).
\end{equation}
 For the FMR  accumulated phase  of Eq.~\eqref{eq:def:phi}, using
  Eqs.~\eqref{eq:alphanbetan:prl}  the   $\alpha_n$
functions appearing above reduce to the $J_n(\Omega_x)$
tfirst  order  Bessel  functions, and the  $\beta_n$  functions  to
combination  of  those  functions.\\
\indent At the first  order perturbation, the synthetic  field
becomes
 \begin{equation}
  \label{eq:h0:prl}
  \vec{h}^{(1)}_s = \Omega_y J_p(\Omega_x)
  \begin{pmatrix}
    0 \\
    \cos(\Phi) \frac{1+(-1)^p}{2} \\
    \sin(\Phi) \frac{1-(-1)^p}{2} \\
  \end{pmatrix}.
\end{equation}
For $p$-odd case the $z$ component  only is different from zero, while
for    $p$    even    this    applies   to    the    $y$    component.
Fig.~\ref{fig:hzplots} (a) reports the $\Omega_x$ dependencies
for  the  $p=1$  $h^{(1)}_{s,z}$  and  $p=2$  $h^{(1)}_{s,y}$ 
components.  The sign changes produced the Bessel functions are
verified in the Cs experiment of~\cite{Bevilacqua_PRL_20}. \\
\indent  The  $Q_x$ amplitude  of  the  second order  synthetic  field
$\vec{h}^{(2)}_{s,x}$ component of Eq.~\eqref{eq:h20:synte} is derived
in  Eq.\eqref{eq:Qxx:prl}  for  the  cosine drivings  and  plotted  vs
$\Omega_x$ for  $(p=1,2)$ in Fig.~\ref{fig:hzplots}(b). This  second order
contribution is  ten times weaker  than the first order  ones.\\
\indent  Outside
the     perturbation    regime,     the    numerical     analyses for $p=1$ produce the 
$h_{z}$ effective values of the Fig.~\ref{fig:hzplots}(c) 2D
$(\Omega_x,\Omega_y)$  plot. Notice that the numerical analysis does not provide the synthetic field information associated to the perturbation treatment. That figure evidences the periodic structures in the spin dynamic response to the dual dressing. In this strong regime the $h_x$ value is comparable to $h_z$, while $h_y$ remains identically zero.\\ 
\indent The first order g-tensor results
\begin{equation}
  \label{eq:gtensor:prl}
  \ten{g}^{(1)} =
  \begin{pmatrix}
    1 & 0 & 0 \\
    0 & J_0(\Omega_x) &0\\
    0 & 0 & J_0(\Omega_x)\\
  \end{pmatrix}.
\end{equation}
\indent  This  diagonal $\ten{g}^{(1)}$,  as  in  the original  single
dressing treatment of ref.~\cite{LandreCohen_70}, denotes that the tensor  principal axes are parallel to the coordinate  axes.   Numerical results  for $J_0$ Bessel  dependence of
the     $\ten{g}^{(1)}$     on     $\Omega_x$     are     shown     in
Fig.~\ref{fig:gplots}(a).  \\ 
\indent  The  $\alpha_n$  and  $\beta_n$ dependencies  on  the  Bessel
functions of Eqs.\eqref{eq:Qij:prl}  determine the $Q_{ij}$ amplitudes
of    the   $\ten{g}^{(2)}$    tensor    components  of
Eq.~\eqref{eq:g2ord}.   For the  $p=1$  harmonic  the four  components
different from zero are 
plotted  vs $\Omega_x$  in Fig.~\ref{fig:gplots}(b),  all of  them ten
times  weaker  than the  first  order.   The  plots evidence  the  non-symmetric  form  of  the  tensor.   The  $\ten{g}^{(1)}+\ten{g}^{(2)}$
tensor  three  principal values  are either  all real  or one  real and  two
complex  conjugates, as  presented in  Figs.~\ref{fig:gplots}(c) and (d). While
for   $(p=1,\Phi=0)$    the   three   values   are    all   real,   the
$(\Omega_x,\Omega_y)$  region of  complex values  increases with
$\Phi$. For $(p=1,\Phi=\pi/2)$ the regions with two complex values
are denoted in red in the 2D plot of Fig.~\ref{fig:gplots}(d).  Notice
that,  from the  physical  point of  view, the  spin  response should  be
symmetric  by swapping  $\Omega_x$ and  $\Omega_y$. This symmetry does  not appear in the Fig.~\ref{fig:gplots}(d)
plot, because for each $\phi$ phase   the gauge transformation to  the
FMR reference frame leads to a different eigenvalue phase shift. \\ 
\begin{figure}
   \centering
    \includegraphics [angle=0, width= 0.6\columnwidth] {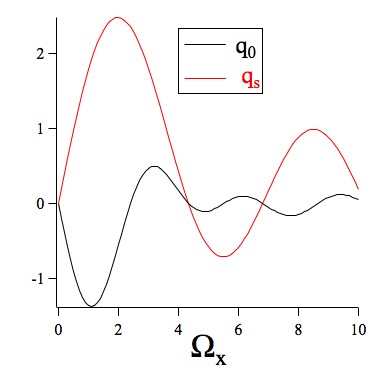}
        \caption{Results for the $q_0$ and $q_s$ functions determining
          the  $\ten{f}^{(2)}$ tensor  components vs  $\Omega_x$. They
          are independent of the $\Omega_y$ parameter.} 
 \label{fig:f2plots}
\end{figure}
\indent Fig.~\ref{fig:f2plots} reports the  $q_0$ and $q_s$ quantities
of  $\ten{f}^{(2)}$ components  as derived in
Eqs.~\eqref{eq:qi:prl}, $q_c$ being equal zero. At  this perturbation  level no  dependence on
$\Omega_y$  is  present.  The  nonlinear Zeeman  contribution  to  the   effective   field   assumes  a   maximum   value   $\approx
0.5(\omega_{0y}^2+\omega_{0z}^2)$ for both diagonal and non-diagonal components of the tensor. 

\subsubsection{Shifted cosine driving}
\label{sec:shift:cos}
In order to present the role played  by  the
initial phase  $\psi$ within the  FMR gauge  transformation. We examine the case of the dual dressing  with
\begin{eqnarray}
  \label{eq:sd:not:even}
  s_x(\tau) &=& \cos(\tau + \psi), \nonumber \\
  s_y(\tau)&=&\cos\left(p(\tau + \psi )+\Phi\right),
\end{eqnarray} 
with  again  $\Phi$  the  relative phase  between  the  drivings.   An
equivalent role of the initial phase  occurs for the Floquet gauges of
periodically  driven   optical  lattices~\cite{GoldmanCooper_15}.   In
addiion,  the $\Omega_y=0$  case with  the $\psi$  phase not  properly
controlled describes all the single dressing atom experiments of
refs.~\cite{HarocheCohen_70_1,HarocheCohen_70_2,LandreCohen_70,Muskat1987,ItoYabuzaki_03,EslerTorgeson_07,BeaufilsGorceix_08,Chu_11,Bevilacqua_PRA_12}.

   The    first   order    synthetic   field   is    given   by
Eq.~\eqref{eq:h_synt:1o} with the following $\beta_0$ expression
\begin{equation}
  \label{eq:beta0:shift}
  \beta_0 =
  \begin{cases}
    J_p(\Omega_x) \cos( \Phi) \e^{-i\,\xi\, \sin(\psi)} \qquad p \; \mathrm{even}\\
    i J_p(\Omega_x) \sin( \Phi) \e^{-i\,\xi\, \sin(\psi)} \qquad p \; \mathrm{odd}
  \end{cases}.
\end{equation}
For the    
$\ten{g}^{(1)}$ tensor we obtain
\begin{equation}
  \label{eq:G1:shift}
  \ten{g}^{(1)} 
  =
  \begin{pmatrix}
    1 & 0 & 0 \\
    0 & J_0(\Omega_x) & 0 \\
    0 & 0 & J_0(\Omega_x)
  \end{pmatrix}
  \begin{pmatrix}
    1 &0 &0 \\
    0 & \cos(\tilde{\Psi})  & -\sin(\tilde{\Psi}) \\
    0 & \sin(\tilde{\Psi})  & \phantom{-}\cos(\tilde{\Psi})
  \end{pmatrix}
\end{equation}
where $\tilde{\Psi}=\Omega_x \sin(\psi)$ and  the tensor is written
as   the  product   of  a contraction  and  a rotation.   The
non-diagonal matrix elements depending on the $\psi$ parameter lead to
complex principal values  corresponding to  phase rotations in the plane
orthogonal to  the single dressing field. As presented in the following Section the $\psi$ phase  modifies the
detected spin evolution .

\section{Spin evolution and detection}
\label{sec:spindetection}
\subsection{Micromotion}
The dressing operation modifies mean value and time evolution of  the spin components. These quantities are derived from the $U(\tau)$ time evolution operator of Eq.~\eqref{eq:U:ri} rewritten using  Eq.~\eqref{eq:form:U:floquet}  as 
\begin{equation}
U(\tau) =   \e^{-i \varphi_x  \sigma_x/2} \, e^{-i{\cal K}(\tau)} \; \e^{-i \Lambda \; \tau}.
\label{eq:Umicrom}
\end{equation}
leading to the  micromotion operator in the lab  frame
\begin{equation}
{\cal M}(\tau)=\e^{-i \varphi_x(\tau)  \sigma_x/2}\e^{-i {\cal K}(\tau)}.
\label{eq:Mmicrom}
\end{equation}
\indent The ${\cal K}$ kick operator, as derived at the first order perturbation in Appendix~\ref{kickoperator}, is the sum of two terms, one proportional to $\vec{\omega}_0$, and the other to $\Omega_y$. Therefore  for all the single and dual dressing experiments performed so far with low values for those fields, the spin micromotion of  Eq.~\eqref{eq:Mmicrom} is dominated by the first term determined by the transformation to the FMR frame.  
 
\subsection{Expectation values}
 Using  the  Pauli  matrix   exponentiation  and  the  effective  field
$\Lambda$    expression    of    Eq.~\eqref{eq:effectivefield2},    the
$\sigma_x(\tau)$ operator becomes 
\begin{eqnarray}
    &\sigma_x(\tau)= U(\tau)^{\dagger} \sigma_x(0) U(\tau)\approx \left[\left(1-u_x^2\right)\cos\Omega_L\tau+u_x^2\right] \sigma_x(0) \nonumber \\
    &+ \left[u_xu_y\left(1-\cos  \Omega_L\tau \right)- u_z\sin(\Omega_L\tau)\right] \sigma_y(0)\nonumber  \\
    &+\left[u_xu_z\left(1-\cos  \Omega_L\tau\right)+u_y\sin\Omega_L\tau)\right]\sigma_z(0),
    \label{eq:sx:t:gen}
\end{eqnarray}
where the kick operator contribution is
neglected.  A   similar  algebra  derives  the   $\sigma_y(\tau)$  and
$\sigma_z(\tau)$ evolutions.
 For an initial $\langle\sigma_x(0)\rangle=1$ eigenstate, the spin expectation
 values, measured by the absorption or dispersion of lasers propagating along the Fig.~\ref{fig:setup} axis, are 
\begin{align}
    \langle \sigma_x(\tau) \rangle  & = 
       (1 - u_x^2) \cos ( \Omega_{\mathrm{L}} \tau) + u_x^2,  \nonumber \\
  \langle \sigma_y(\tau) \rangle  & = 
    \left[    u_y   \sin\varphi_x+  u_z    \cos\varphi_x    
    \right]\sin(\Omega_{\mathrm{L}} \tau) + \nonumber \\
    &\phantom{=}
    \left[u_x u_y \cos \varphi_x  - u_x u_z \sin\varphi_x \right]
    (1- \cos ( \Omega_{\mathrm{L}} \tau )), 
             \nonumber \\
              \langle \sigma_z(\tau) \rangle &= 
    \left[    u_z    \sin\varphi_x    -   u_y    \cos\varphi_x
    \right]\sin(\Omega_{\mathrm{L}} \tau ) +\nonumber \\
    &\phantom{=}
    \left[ u_x u_z \cos\varphi_x + u_x u_y \sin\varphi_x \right]
    (1- \cos ( \Omega_{\mathrm{L}} \tau )),
\label{eq:sxyz:t}
\end{align}
where $\varphi_x(\tau)$ is given by Eq.~\eqref{eq:sd:prl}. These expressions generalize the derivation in ref.~\cite{GolubLamoreaux_94}. The  spin coherences  contain two separate  time dependencies,
the $\Omega_L$ frequency precession and  a more complex micromotion one determined
by   the  $\cos   (\varphi_x(\tau))$   and  $\sin   (\varphi_x(\tau))$
functions at the harmonics of the driving frequency.   

\subsection{Detection in shifted cosine single dressing}
\label{subsec:shifteddetection}
The  spin  detection in  the  pioneer  single dressing  experiment  by
Landr\'e   et  al.~\cite{LandreCohen_70}   highlights the role  of the off-diagonal elements in
the $\vec{g}$  tensor of Eq.~\eqref{eq:G1:shift}.  No
phase control is  applied to the  dressing field, and that  experiment is
described    by   the    shifted  single  cosine    treatment   of
Sec.~\ref{sec:shift:cos} with  $\Omega_y=0$.  With the
$\langle \sigma_x(0)\rangle=1$  mercury  atoms  prepared initially, at  $t=0$ time a weak  $\omega_{0y}\ne 0$  magnetic field  is
switched on. No  synthetic field is created because $\Omega_y=0$. The    effective   field    $\vec{h}$   vector   derived    from
Eq.~\eqref{eq:G1:shift} is 
\begin{equation}
  \label{eq:h:Landre}
  \vec{h} = 
  \begin{pmatrix}
    0\\
    \omega_{0y} J_0(\Omega_x)\cos(\Omega_x\sin(\psi)) \\
    \omega_{0y} J_0(\Omega_x)\sin(\Omega_x\sin(\psi))
  \end{pmatrix},
\end{equation}
leading to $\Omega_L=\omega_{0y}|J_0(\Omega_x)|$ and to   
the $\vectorsym{u}$  orientation vector 
\begin{equation}
  \label{eq:u:Landre}
  \vec{u} = 
  \begin{pmatrix}
    0\\
    \sgn(J_0)\cos(\Omega_x\sin(\psi)) \\
    \sgn(J_0)\sin(\Omega_x\sin(\psi))
  \end{pmatrix},
\end{equation}
with no dependence on the static field. Using     Eqs.~\eqref{eq:coeff:prl} 
for     dealing     with     the
$\cos(\varphi_x(\tau))$ and  $\sin(\varphi_x(\tau))$  of Eqs.~\eqref{eq:sxyz:t},  we
obtain 
\begin{eqnarray}
  \langle\sigma_x(\tau) \rangle &=& \cos( \Omega_L \tau), \nonumber \\
  \langle\sigma_y(\tau) \rangle &=&\sin( \Omega_L \tau) \sin\left[\Omega_x \sin(\tau + \psi)\right] \nonumber \\ 
                             &=& 2 \sin( \Omega_L \tau)
                                 \left[ J_1 \sin( \tau + \psi) +
                                 J_3  \sin( 3\tau + 3\psi) +
                      \ldots \right],  \nonumber \\
  \langle \sigma_z(\tau)  \rangle &=& -\sin( \Omega_L \tau)\cos\left[ \Omega_x
                                      \sin(\tau+ \psi)\right] \nonumber \\
  &=&  -\sin( \Omega_L  \tau)
                     \left[ J_0 + 2 J_2
                     \cos( 2\tau + 2 \psi) +
                     \ldots \right]. 
\end{eqnarray} 
The $J_0$ response  in $\langle \sigma_z(\tau)  \rangle $ is 
produced  by  the $\cos  (\varphi_x(\tau))$  dependence,  as pointed  out
in~\cite{BeaufilsGorceix_08}.  In the  Landr\'e et al~\cite{LandreCohen_70} experiment the
oscillations at  the frequency $\Omega_L$ of  the $\langle\sigma_x(\tau) \rangle$ and  $\langle\sigma_z(\tau) \rangle$
spin components are detected by probe beamd propagating along those axes of Fig.~\ref{fig:setup}.  The  $xz$ plane spin evolution 
follows an ellipse contracted by  $J_0(\Omega_x)$ on one axis.  Within
that $\Omega_L$ detection, the  $\vec{g}$
tensor non-diagonal form of Eq.~\eqref{eq:G1:shift} does not play any role. In  a detection sensitive
to  the sideband  frequencies, the $\psi$ dependence of the non-diagonal tensor terms, and also the micromotion contributions, can be detected.  For  an arbitrary static field orthogonal
to the  dressing $x$  axis, an  ellipse evolution  takes place  on the
plane perpendicular  to the static  field. Such evolution  matches the
uniaxial cylindrical  symmetry associated to
the single dressing. 

\begin{figure}[!!b]
   \centering
    \includegraphics [angle=0, width= 0.8\columnwidth] {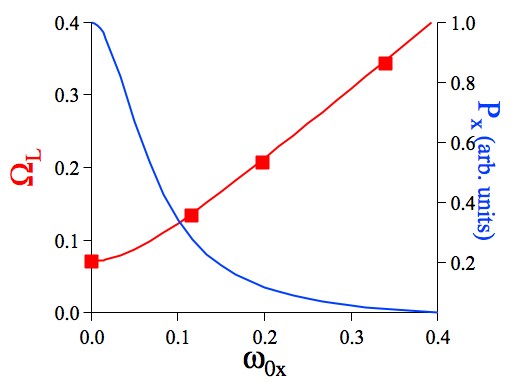}
        \caption{The $\Omega_L$   Larmor frequency   (left  axis)   and  the  $P_{S_x}$ peak of the $\Omega_L$ oscillating  Faraday signal peak (right axis)  vs the applied
          $\omega_{0x}$ magnetic field measured in  the Cs dual dressing experiment~\cite{Bevilacqua_PRL_20}. Parameters:  $\omega_{0z}=0.1993(2)$, $\Omega_x=1.833(5)$, 
          $p=1$,    $\Omega_y=0.0118(1)$,    and   $\Phi=\pi/2$. } 
 \label{fig:CsB0x}
\end{figure}

\subsection{Detection in dual dressing}
In  ref.~\cite{Bevilacqua_PRL_20} the  dual dressing is tested for optically pumped Cs atoms.  Because the detection is synchronized
by the  $x$ dressing field and the $\Omega_x$ phase is fixed in the experiment, the  system is described by  the sinusoidal
laws of  Eq.~\eqref{eq:sinusoidal}.  Optical pumping  along the $x$
axis  and synchronous  with  the $\Omega_L$  precession frequency  is
applied  to the  spins.   Faraday detection of a probe beam propagating on the $x$ axis of Fig.~\ref{fig:setup} monitors the   $\langle\sigma_x(\tau) \rangle$
time response at that frequency. Therefore the $\Omega_L$ dressed Larmor frequency and the $P_x$ peak amplitude of a signal  proportional to the $\Omega_L$ component of  $\langle \sigma_x(\tau)  \rangle$ are measured. We examine here the $p=1$  dual dressing  detection at fixed values  of $(\Omega_x,\Omega_y)$
and $\omega_{0z}$,  as a  function of  an $\omega_{0x}$ applied field. At the first perturbation order this field
plays a dual role. It modifies  $\Omega_L$,  given  by 
\begin{equation}
 \label{eq:OmL:Cs}
  \Omega_L^{(1)}(\omega_{0x}) = 
\sqrt{\omega_{0x}^2+ 
    \left(J_0(\Omega_x)\omega_{0z}+ 
     J_1(\Omega_x) \Omega_y \sin (\Phi)\right)^2},
\end{equation}
and the $\vec{u}$ spatial  orientation, given by 
\begin{equation}
  \label{eq:h:Cs}
  \vec{u} = 
  \frac{1}{\Omega_L^{(1)}}\begin{pmatrix}
    \omega_{0x}\\
   0
    \\
    J_0(\Omega_x)\omega_{0z}+ 
     J_1(\Omega_x) \Omega_y \sin (\Phi)
  \end{pmatrix}.
\end{equation} 
The Faraday detection  signal monitors the following $\langle \sigma_x(\tau)  \rangle$  derived from Eq.\eqref{eq:sxyz:t}: 
\begin{equation}
 \langle \sigma_x(\tau)  \rangle=A_x\cos(\Omega_L \tau)+\left(\frac{\omega_{0x}}{\Omega_L}\right)^2,
\end{equation}
with the $A_x$ oscillation amplitude 
\begin{equation}
A_x= \left[1-\left(\frac{\omega_{0x}}{\Omega_L}\right)^2\right].
\end{equation}
The measured $\Omega_L$ Larmor  frequency and $P_x$ value proportional
to  $A_x$   are  plotted   in  Fig.~\ref{fig:CsB0x}  vs   the  applied
$\omega_{0x}$  field, for  the experimental  parameters in  the figure
caption.    The    $\Omega_L$   dependence    is   well    fitted   by
Eq.~\eqref{eq:OmL:Cs}.  The  Lorentzian shaped decrease of  $P_x$ with
$\omega_{0x}$, observed in the  experiment but not carefully measured,
is produced by  the change in the spin precession  plane orthogonal to
the $\vec{u}$  orientation vector as modified  by $\omega_{0x}$. These
results demonstrate  the triaxial symmetry  of the spin  response. The
micromotion   does   not   appear   on   the   experimental   observed
$\langle \sigma_x(t) \rangle$  signal. Eqs.\eqref{eq:sxyz:t} show that
it can be monitored detecting the orthogonal spin components.

\section{Dual dressing applications}
The control provided by the  dual dressing leads to quantum technology
advances in  a variety of experimental  configurations. Different ones
are presented  here, based on  the flexible effective field  tuning in
amplitude and direction, for either one or two different spins.

 \subsection{Increased magnetic response}
\label{sec:acceleration}
\indent  Because   in  magnetic  resonance  and   other  spectroscopic
techniques  the   detection  sensitivity   increases  with   the  spin
precession frequency,  it is  important to increase,  at a  given real
static magnetic  field, the $\Omega_L$ Larmor  frequency. This result,
implemented  by  the  dual  dressing, is  measured  by  the  following
$a\Omega_L$ accelerated Larmor frequency:
\begin{equation}
 a\Omega_L=\frac{\Omega_L}{\omega_0},
 \label{eq:acceler}
 \end{equation}
 \begin{figure}
   \centering
   \includegraphics [angle=0, width= 0.8\columnwidth] {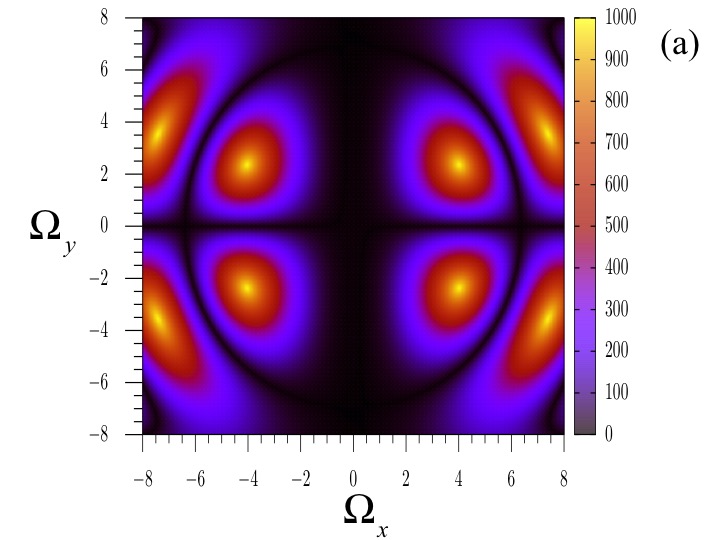}
   \includegraphics [angle=0, width= 0.6\columnwidth] {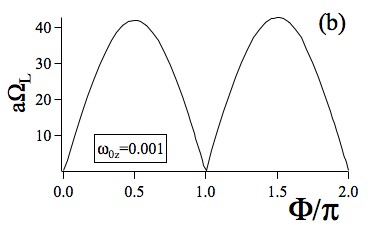}
    \caption{In (a) numerical results for  $a\Omega_L$        in    the     2D $(\Omega_x,\Omega_y)$   plane  at   $\omega_{0z}=0.001$, $\omega_{0x}=\omega_{0y}=0$, $p=3$, and
      $\Phi=\pi/2$.  The  $\simeq$1000   maximum  value  is  reached  at
      $(\Omega_x\simeq\pm3.9,\Omega_y\simeq\pm2.4)$                          and
      $(\Omega_x\simeq\pm7.5,\Omega_y\simeq\pm3.6)$  values. In  (b)  from   the
      perturbation    treatment,  $a\Omega_L$  vs   $\Phi$   derived     at   $p=3$,    $\omega_{0z}=0.001$,
      $\Omega_x=3.9,\Omega_y=1$, with maximum $\approx 43$.}   
  \label{fig.wz_0.001}
\end{figure}
equal to one in the absence of the dressing. Owing to the periodic structure of the
eigenfrequency  Brillouin zones, the maximum allowed $\Omega_L$ value is one. Therefore  in the low magnetic field  range of our
 interest  $a\Omega_L$   becomes  quite
 large.  As  shown   in  the numerical data of Fig.~\ref{fig.wz_0.001}(a),  for  $p=3$,
 $\omega_{0z}=0.001$ and  a proper  choice of  the $(\Omega_x,\Omega_y)$
 values, $a\Omega_L$ reaches a maximum value around one thousand
 as allowed by  the Brillouin  zone boundary.  Such  high frequency response leads  to a
 very high  sensitivity in the  spin detection.  Brillouin   zone  boundary  ratios  are   reached for all the $\omega_{0z}$ values.  A  $p=3$
 perturbation   treatment   produces   the  $a\Omega_L$   of
 Fig.~\ref{fig.wz_0.001}(b)  with   a  maximum  value  at
 $\Phi=\pi/2$. The $\Phi$ phase
      dependence  of the  perturbation treatment  applies also  to the
      strong dressing regime. For each phase the maximum is obtained
 at different dressing parameters.  Similar periodic maxima appear also for the $p=1$, $\Omega_x=\Omega_y$ and $\Phi=\pm\pi/2$ rotating dressing case where the Larmor frequency is determined by the  effective field in the rotating frame. They are originated by the folding of the dressed Larmor frequency into the periodic Brillouin structure.\\
 \begin{figure}[!!t]
   \centering
   \includegraphics [angle=0, width= 0.7\columnwidth] {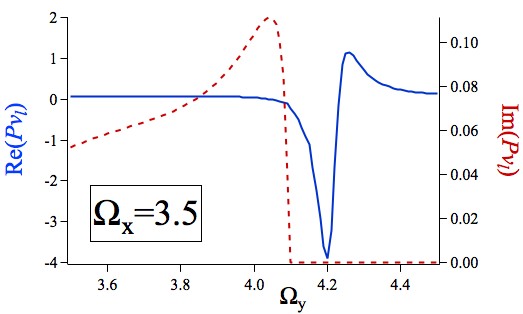}
    \caption{Numerical results for $Pv_l$, the largest principal value of the $\vec{g}$ tensor, vs  $\Omega_y$, for $\Omega_x=3.5$, $p=1$ and $\Phi=\pi/2$. Real Re($Pv_l$) and complex Im($Pv_l$) parts are plotted. The  $\Omega_y$ parameter is varied in the transition region from one real and two complex principal values to  three real values.  Notice that the largest absolute Re($Pv_l$) value is negative. }   
  \label{fig:gIncrease}
\end{figure} 
\indent In atomic  interferometry experiments with a  Stern-Gerlach deflection, a magnetic  field gradient splits the  particles into spatially
 separated  paths, for instance see~\cite{MargalitFolman_15}. The  accelerated dressed Larmor frequency may be used to increase that deflection owing to the modified linear Zeeman splitting. This occurs when the bare Land\'e g-factor, equal one in our units,  is replaced by a large principal value of the $\vec{g}$-tensor.  Such case is presented in Fig.~\ref{fig:gIncrease} for the caption parameters. The absolute principal value of $Pv_l$ (the largest one) increases  for dressing parameters close to a transition from one real and two complex conjugates to three real ones, as those plotted in Fig.~\ref{fig:gplots}(c).  The four times increase produced by the  dual-dressing increase  is certainly useful in  experiments as that quoted above. The effective  negative sign of the effective Land\'e g-factor should be no problem for the experimentalists.   In an experimental implementation, because the remaining $\vec{g}$-tensor principal values remain around one or below, the dressing fields should be oriented in space in order to align the principal axis of interest with the experimentally applied Stern-Gerlach magnetic gradient.  \\ 
    
\subsection{Magnetic field compensation}
\label{sec:compensation}
We target here the compensation of a static field arbitrarily oriented in space by  generating an opposite sign synthetic magnetic field.   3D compensation is expressed by the following expression:
\begin{equation}
\Omega_L=0,
\label{eq:def:compensation}
\end{equation}
or equivalently
\begin{equation}
h_x=h_y=h_z=0.
\label{eq:def:compensation2}
\end{equation}
Instead for 2D or 1D we impose $h_i=0$ for the required dimensions. Compensations of 3D and 2D magnetic fields are required in high resolution experiments. For instance,  in interferometric investigations  with artificial
 or  natural  atoms  as in~\cite{OnoNori_19,AmitFolman_19},  the  fine
 tuning of  the magnetic response with  a controlled  compensation on
 different  spatial   directions  produces a higher precision. \\
\begin{figure}
   \centering
      \includegraphics [angle=0, width= 0.9\columnwidth] {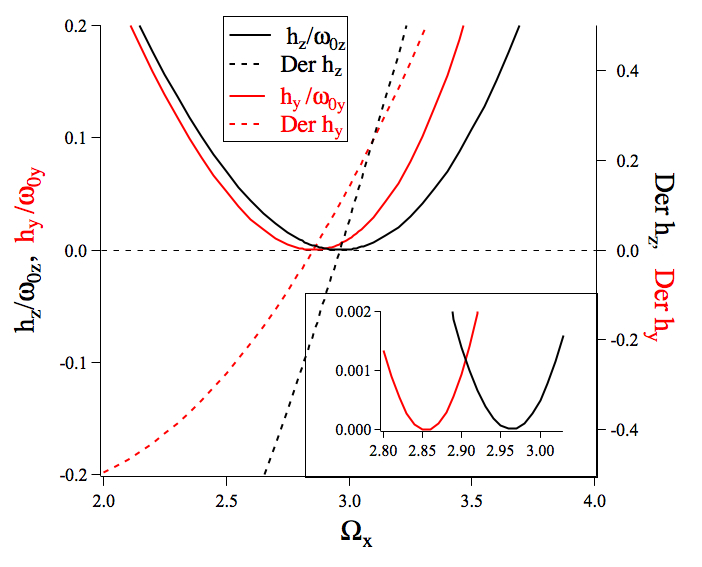}
        \caption{On the left scale, compensation results for the $h_y,h_z$ effective  transverse magnetic fields vs the strong $\Omega_x$ dressing field in a 
        $\omega_{0x}=0.1$, $\omega_{0y}=\omega_{0z}=1\cdot10^{-4}$   configuration, with weak dressing fields
 $\Omega_{z5}=0.6\cdot10^{-5}$, $\Omega_{y6}=2.4\cdot10^{-5}$. The plot of the normalized $h_y/\omega_{0y},h_z/\omega_{0z}$ ratios evidences the magic compensations. On the right scale the $\partial h_y/\partial \Omega_x,\partial h_z/\partial \Omega_x$ derivatives vs $\Omega_x$ in the 
        second-order magic compensation search. Even if the null derivatives are not exactly coincident for the magic compensation, the sensitivity to the strong dressing fluctuations is reduced by three orders of magnitude. This appears in the inset plot  for the normalized effective fields in the  $\Omega_d$   (2.87,2.92) range.  } 
 \label{fig:CompResults}
\end{figure}
\indent Nearly periodic zero values of the Larmor frequency appear in Figs.~\ref{fig:eigenvalues}(a) and~ \ref{fig.wz_0.001}(a). From the $\vec{h}^{(1)}_s,\vec{g}^{(1)}$ expressions of Eqs.~\eqref{eq:h0:prl} and~\eqref{eq:gtensor:prl} we derive that the applications of dressing fields along the three spatial directions produce the compensation of an arbitrary magnetic field configuration. However the action of several strong dressings  cannot be  handled by a perturbation approach, and numerical solutions are required. \\
\indent  A  more  ambitious  goal, denoted  as  second  order  magic
compensation or magic shield,  is to  produce a  reduced sensitivity  to fluctuations of the  dressing  field               or      of the          magnetic               field, respectively, ~\cite{ZanonArimondo_12,JonesMartin_13,SarkanyFortagh_14,KazakovSchumm_15,BoothSaffman_18,TrypogeorgosSpielman_18,AharonRetzker_19}. These
magic solutions are  obtained by solving Eqs. \eqref{eq:def:compensation2}, and
simultaneously imposing zero values for their derivatives with respect
to the compensation parameter/s. Second-order magic dressing requires a
nonlinear dependence  on the control parameters,  for instance through
the  Bessel functions  for  the dressing  Rabi frequencies.  Nonlinear
dependencies  on the  static magnetic  fields appear  at the second and higher  order
perturbation treatments.\\ 
\indent We derive  here the dressing parameters realizing a 2D magic compensation for the $^{87}$Rb atomic chip studies of refs.~\cite{LacrouteRosenbusch_10,DeutschReichelRosenbusch2010}  for an applied  300 $\mu$T field. There the target was to reduce the transverse fields from the 0.3 $\mu$T range  into the nT one. For such 2D compensation,   an  
$\Omega_x$  dressing   field is applied parallel to the non-compensated $\omega_{0x}$  field.  Within  the   first-order  perturbation
treatment  we derive  that an  $h^{(1)}_z=0$ effective synthetic field is produced  by a  properly
chosen  $\Omega_{y5}\cos(5\tau+\pi/2)$ driving.  Instead $h^{(1)}_y=0$
is reached adding a properly chosen $\Omega_{y6}\cos(6\tau)$ dressing,
these high harmonics being useful for the $\Omega_x$ magic compensation. Up to the second order perturbation level the 2D compensation is given by
\begin{align}
h_y=&J_6(\Omega_x)\Omega_{y6}+J_0(\Omega_x)\omega_{0y}\nonumber \\
&+\frac{1}{2}Q_{yx}(\Omega_x)\Omega_{y6}\omega_{0x}+2q_s(\Omega_x)\omega_{0x}\omega_{0y}=0,\nonumber \\
h_{z}=   &J_5(\Omega_x)\Omega_{y5}+J_0(\Omega_x)\omega_{0z}\nonumber \\
&+\frac{1}{2}Q_{zx}(\Omega_x)\Omega_{y5}\omega_{0x}+2q_s(\Omega_x)\omega_{0x}\omega_{0z}=0.
\end{align}
\indent The 
second order contributions to these effective fields are greatly reduced
by choosing a large dressing frequency, for instance 30 times
greater  than  the  bare  Larmor   frequency, i.e.,  $\omega_{0x}=0.1$.  The above equations  contain a nonlinear dependence  on the $\Omega_x$
dressing  field and  therefore  allow a  magic  compensation for  that
parameter, as 
presented in  Fig.~\ref{fig:CompResults}.  The above equations not
containing a nonlinear dependence on the $(\omega_{0y},\omega_{0z})$ parameters,  don't support transverse field magic fluctuation shields. For those parameters as well for the $(\Omega_{y5},\Omega_{y6})$ ones,  magic compensations can be determined by numerical analyses. Multiple harmonic  driving  produces interesting compensation schemes to be explored. \\ 

\subsection{Inhomogeneous dressing}
\indent  A spatial  gradient of  the dressing
 field  can  increase the  forces  on  the spins  using  an       inhomogeneous       oscillating
 field as     
  for trapped ions in~\cite{OspelkausWineland_08,WoelkWunderlich_17} and as for the Cs experiments \cite{biancalana_prappl_19,Bevilacqua_PRL_20} where we tested the dual dressing. Therefore the 
 magnetometry  applications   with  $\Omega_{\mathrm{L}}$  
 deliberately spatially dependent  can be enlarged by the dual dressing configuration. Such spatial  distribution is more easily realized by operating on the position dependence of a weak field tuning the dressing of the strong one. 
 The atomic trapping  with spatially inhomogeneous radiofrequency potentials, reviewed in~\cite{PerrinGarraway_17}, can produce new spatial configuration by  replacing the single dressing by the dual one.    \\

 \section{Conclusion and Outlook}
 For a spin one-half system in presence of an arbitrary static magnetic field, we study the dual dressing by a primary field and a secondary one oscillating at a harmonic frequency. Within a perturbative treatment the secondary field acts as a tuning of the first strong dressing. The two fields play an equivalent role in numerical analyses. The dual dressing introduces a very rich dynamics into a quantum system. The standard  spin Larmor precession around the external static field is replaced by a dressed Larmor precession, whose frequency is controlled by the dressing fields. We present dressing parameters where that frequency is, under proper conditions, either increased  one thousand times or  decreased down to zero. These conditions are reached in experiments operating with nT static fields, applying electromagnetic fields whose frequency, amplitude and phase are accessible experimentally. In magnetic resonance, i.e., the dressing by a rotating field, a resonant field creates in the rotating frame a spin precession around the rotating field with frequency determined by the rotating field amplitude. The dual dressing extends that feature to an arbitrary spatial orientation of the spin, and also leads to a precession frequency as large as the electromagnetic field frequency. In addition our non-resonant dressing acts not only on the resonant species, but on all spin species of the sample, as for the investigated two-species case. \\     
 \indent The generalization of the dressed atom to the dual driving configuration
 enriches  the spin control,   produces  features
 useful to several  quantum control directions and  enables new quantum
 technology  explorations.  The  effective/fictitious fields, providing  a simple  and
 direct access to the qubit  control, represent
 the  dual-dressing handle. The commensurable and low harmonic
 driving is a  key component of its great  impact on the spin dynamics. For an initially symmetric spin system, the triaxial
 response  created  by  the  dual  dressing  introduces  a  controlled
 anisotropy, where the effective field orientation is the anisotropy helm.  These characteristics apply also to a multiple harmonic dressing, as from
 a  straightforward  extension  of   our  theoretical  treatment.   \\ 
 \indent As for other modulated systems, the micromotion represents an essential component of the spin dynamics under dual dressing driving. Its main component is associated to the FMR frame transformation, with an additional component produced by the kick operator.  The influence of the kick operator can be neglected operating at low static field, as for all dressing experiments so far. The micromotion produces additional high harmonic components, easily separated by performing a spectral analysis of the detected experimental signal. This spectral separation  allows an employ of the micromotion components to enhance  the fidelities of  quantum gates based on dual-dressed qubits, as performed for the trapped ions qubits, see~\cite{RatcliffeHope2020} and references therein.\\
 \indent We have presented several dual dressing applications in spectroscopy, atomic physics, quantum simulation and computation. For spectroscopy, the controlled increase of the Larmor frequency shifts the spin detection towards higher frequencies, where the experimental sensitivity is larger. On the opposite direction, in atomic physics experiments as atom chips or atomic interferometers the externally applied magnetic field should be compensated in order to improve the experimental precision.  Under the combined action of even and odd harmonics in the secondary dressing field, the synthetic field reduces drastically the magnetic splittings in all chosen directions.\\
  \indent  For  spintronics, the  dual-dressing  can be applied to  artificial  atoms, where  the energy  splittings produced by the solid host  are equivalent to
 static magnetic fields here considered. Our scheme leads to
 a control of those energy  splitting, to be increased  or compensated. A wide range of  g-factor anisotropies appears
 in   solid   state  physics,   from   less   than  one   percent   in
 nitrogen-vacancy    colour    centres    for    precision    magnetic
 sensors~\cite{Doherty_13,RondinJacques_14,SchirhaglDegen_14},  up  to
 thirty  percent  in  InSb   quantum  wells  for  topological  quantum
 computing~\cite{QuKouwenhoven_16}.  The bichromatic  dressing can  be
 used either to compensate the  Land\'e g-factor anisotropy for higher
 precision  magnetic  sensors,  or  to  increase  the  anisotropy  and
 therefore the topological importance of the material.\\ 
\indent The results of our perturbative and numerical treatments evidence the presence of symmetries for the
 eigenvalues  and the  effective/synthetic fields.  These  symmetries
 depend on harmonic order and dressing phase. In~\cite{NeufeldCohen_19} the group
 theory  of  the  dynamical  symmetries in  periodic  Floquet  systems
 is  applied     to  the
 nonlinear   harmonic   generation.    Those  group  operations,  as  rotations,  reflection  and
 different symmetries, as inversion, spatio-temporal  or spatial only,  should be
 used to analyze the response of a dual-dressed spin for a wider parameter range. \\ 
 \indent Owing to the easy experimental implementation  of the double dressing, the introduction  of
 another handle,  as the time dependence  of the dressing
 field   amplitudes, with properly designed  adiabatic  or   superadiabatic  temporal
 evolutions,  can be used to produce  new  quantum  superposition states.  On  a
 different direction,  the application of a  sinusoidal modulation to the dressing fields can
 open new directions  for  the dynamical  driving  of  ultracold
 atoms in optical lattices.
 
 \section{Acknowledgments}
 The authors thank H\'el\`ene Perrin, Andrea Tomadin and Sandro Wimberger for a careful reading and constructive comments of the manuscript.

\appendix
\section{Numerical solution}
\label{sec:num:sol}
The numerical  solution of the  $\Lambda$ Floquet matrix  requires the
propagation of  the operator  $U(\tau)$ from $\tau=0$  to $\tau=2\pi$,
then the  diagonalization of  $U(2\pi) =  \e^{-i \,2\pi  \, \Lambda}$~\cite{BukovPolkovnikov_15},
leading to
\begin{equation}
  \label{eq:diag:U}
  U(2\pi) | \lambda_\pm \rangle = e^{-12\pi\lambda_\pm}| \lambda _\pm \rangle,
\end{equation}
with $\lambda_\pm$ eigenvalues and the Floquet $\Lambda$ matrix 
\begin{equation}
  \label{eq:Lambda:num}
  \Lambda = \sum_{j=\pm}\lambda_j |\lambda_j\rangle\langle \lambda_j |.
\end{equation}
The components of the vector $\vec{h}$ are obtained as
\begin{equation}
  \label{eq:def:F:num}
  h_j =  \mathrm{trace}\left( \Lambda \sigma_j\right), \qquad j=x,y,z.
\end{equation}

\section{FMR Floquet-Magnus expansion}
\label{sec:magn:expa}
The Magnus expansion  writes in an exponential form  the time evolution
operator of  a linear  system~\cite{magnus2009}.  When applied  to the FMR
$U_I$ time evolution operator of Eq.~\eqref{eq:U:ri} under a generic $H_I$ Hamiltonian
\begin{equation}
  \label{eq:U:def}
  i \dot{U_I}(\tau) = H_I(\tau) \; U_I(\tau), 
\end{equation}
it parametrizes the operator as
\begin{equation}
  \label{eq:U:param}
  U_I(\tau) = \e^{-i W(\tau)}.
\end{equation}
This leads for the $W(\tau)$ exponent
\begin{equation}
  \label{eq:W:non-linear}
  \dot{W}(\tau) = H_I(\tau) + \frac{i}{2} [W(\tau), H_I(\tau)]
 + \ldots 
\end{equation}
 If     $H_I(\tau)$     is     ``small'' , i.e., 
$  H_I  \rightarrow \epsilon  H_I$  for  a small $\epsilon$,  expressing
$W = \epsilon W_1  + \epsilon^2 W_2 + \ldots$ one  finds for the first
terms
\begin{eqnarray}
    W_1(\tau) &=& \int_0^\tau H_I(\tau_1)\; \de \tau_1, \nonumber \\
    W_2(\tau) &=& \frac{-i}{2}\int_0^\tau \de \tau_1 \; \left[ H_I(\tau_1), W_1(\tau_1) \right].
    \label{eq:magnus:ord}
\end{eqnarray}
Within the  Fourier description of  Eq.~\eqref{eq:form:U:floquet},  the Magnus  expansion is
applied   to  the   ${\cal  K}$   and  $\Lambda$   exponents.  Letting
$H     \rightarrow    \epsilon     H$    as     above,    we     write
${\cal  K}(\tau)  =  \epsilon  {\cal  K}_1(\tau)  +  \epsilon^2  {\cal
  K}_2(\tau)\ldots                        $                        and
$ \Lambda  = \epsilon  \Lambda_1 + \epsilon^2\Lambda_2  +\ldots$. Thus
the following formulas are obtained:
\begin{eqnarray}
  \label{eq:F1:L1}
    \Lambda_1 &=& \frac{1}{2\pi} \int_0^{2\pi} H_I(\tau) \mathrm{d}\,\tau, \nonumber \\
    {\cal K}_1(\tau) &=& \int_0^{\tau} H_I(\tau') \mathrm{d}\,\tau' -\tau \Lambda_1, \nonumber \\
    \Lambda_2  &=& -\frac{i}{4  \pi} \int_0^{2\pi}  \left[ H_I(\tau)  +
      \Lambda_1,
      {\cal K}_1(\tau) \right] \mathrm{d}\,\tau, \nonumber\\
    {\cal K}_2(\tau)  &=& -  \tau \Lambda_2  -\frac{i}{4\pi} \int_0^{\tau}
    \left[   H_I(\tau')  +   \Lambda_1,   {\cal  K}_1(\tau')   \right]
    \mathrm{d}\,\tau'.
\end{eqnarray}
 
\section{Fourier expansions of the driving}
\label{sec:fourier}
In order to simplify the mathematical derivations, 
we introduce few time-dependent functions.
For  a generic accumulated  phase introduced  by Eq.~\eqref{eq:def:phi},  more general than the sinusoidal one, we
define the functions  $(c_0 = \cos \varphi_x, s_0  = \sin \varphi_x)$,
$(c_1  =  s_y\,  \cos  \varphi_x,s_1  =  s_y\,  \sin  \varphi_x)$  and
$(\alpha_n,\beta_n)$ given by
\begin{eqnarray}
  \label{eq:def:csp:coeffs}
    \e^{i \varphi_x (\tau)} & \equiv& c_0(\tau) + i\, s_0(\tau)
                            =  \sum_{n=-\infty}^{+\infty}   \alpha_n  \e^{  i\,n
                            \tau},  \label{eq:def:c0s0:alphan} \nonumber \\
    s_y(\tau)\e^{i \varphi_x (\tau)} & \equiv& c_1(\tau) + i\, s_1(\tau)
                                       =  \sum_{n=-\infty}   \beta_n  \e^{  i\,n
                                       \tau}
                                     \label{eq:def:c1s1:betan},
\end{eqnarray}
From the $\alpha_n$ definition, $|\alpha_0|  \leq 1$ is easily derived. \\
\indent  In  addition we  introduce  $C_i(\tau)$,  $S_i(\tau)$ as  the
primitives of the lower case functions, respectively, as
\begin{subequations}
\label{eq:def:CSG:coeffs}
\begin{align}
    \int_0^\tau \e^{i \varphi_x (\tau')} \de \tau'
                          & \equiv C_0(\tau) + i\, S_0(\tau) \nonumber \\
                            & = \alpha_0 \tau
                            -i \sum_{n \neq 0}
                            \alpha_n \frac{ \e^{ i\,n\tau} -1 }{n}
                              \label{eq:def:CC0SS0:alphan} \\
    \int_0^\tau s_y(\tau')\e^{i \varphi_x (\tau')} \de \tau'
                          & \equiv C_1(\tau) + i\, S_1(\tau) \nonumber \\ 
                            &= \beta_0 \tau
                            -i \sum_{n \neq 0}
                            \beta_n \frac{ \e^{ i\,n\tau} -1 }{n}.
                              \label{eq:def:CC1SS1:alphan}
\end{align}
\end{subequations}
Explicitly we obtain
\begin{align}
  \label{eq:expl:C0:S0}
    C_0 &= \mathrm{Re}(\alpha_0)\tau 
        \phantom{=}+\sum_{n\geq 1}\left[
          \frac{\mathrm{Im}(\alpha_n - \alpha_{-n})}{n} \left(
          \cos(n\tau) -1
          \right)
          + \right. \nonumber \\ 
        & \phantom{+\sum_{n\geq 1}}\left. \;\;\;\;\;\frac{\mathrm{Re}(\alpha_n + \alpha_{-n})}{n} \sin(n\tau)  \right] \\
    S_0 &= \mathrm{Im}(\alpha_0)\tau
        \phantom{=}+\sum_{n\geq 1}\left[
          \frac{-\mathrm{Re}(\alpha_n + \alpha_{-n})}{n} \left(
          \cos(n\tau) -1
          \right)
          + \right. \nonumber \\ 
        & \phantom{+\sum_{n\geq 1}}\left. \;\;\;\;\;\frac{\mathrm{Im}(\alpha_n - \alpha_{-n})}{n} \sin(n\tau)  \right] 
\end{align}
and similar  expression for  $C_1$ and $S_1$  replacing $\alpha_n$
by $\beta_n$.\\ 
\indent For the  cosine signal drivings of Eq.~\eqref{eq:sinusoidal} and  the sinusoidal accumulated
phase of Eq.~\eqref{eq:def:phi} we obtain
\begin{equation}
  \label{eq:coeff:prl}
  \e^{i\, \varphi_x(\tau) }  = \cos(\varphi_x(\tau))+i\sin(\varphi_x(\tau))=\sum_n J_n(\Omega_x) \e^{i\,n \, \tau},
\end{equation}
where $J_n$ are the Bessel functions of  first kindx.  Therefore the
$\alpha_n$ and $\beta_n$ functions become
 \begin{eqnarray}
 \alpha_n &=& J_n(\Omega_x), \nonumber \\
    \beta_n & =& \frac{1}{2}
  \left(
    \e^{i \Phi} \, J_{n-p}(\Omega_x) + \e^{-i \Phi} \,J_{n+p}(\Omega_x) 
  \right).
    \label{eq:alphanbetan:prl}
\end{eqnarray}
The $c_0$ and $s_0$ functions become
\begin{align}
    \label{eq:sincos:Jacobi:Anger}
    c_0(\tau)&= J_0(\Omega_x)
               + 2 \sum_{n\geq 1}J_{2n}(\Omega_x)
               \cos( 2 n \tau), \\
    s_0(\tau)&= 2  \sum_{n\geq 0}J_{2n+1}(\Omega_x)
               \sin( (2n+1)\tau ), 
\end{align}
and the $c_1$ and $s_1$ functions become
\begin{align}
    \label{eq:c1s1}
  2 c_1(\tau) &= \cos\Phi (J_{-p} + J_p)  + \nonumber \\
              &\sum_{n\geq 1}\left[ - \sin \Phi\,
                \big(J_{n-p}- J_{-n-p}-J_{n+p}+ J_{-n+p}\big)\,
                \sin(n\tau) \right. \nonumber \\
              &\left. + \cos\Phi
                \big( J_{n-p}+ J_{-n-p}+J_{n+p}+ J_{-n+p}\big)\,
                \cos(n \tau)\right],
                \nonumber\\           
  2 s_1(\tau)&= \sin\Phi \,(J_{-p} - J_p)  +\nonumber \\
            & \sum_{n\geq 1}\left[ \cos \Phi
              \big( J_{n-p}- J_{-n-p}+J_{n+p}- J_{-n+p}\big) \,
              \sin(n\tau) \right. \nonumber \\
            &\left. + \sin\Phi
              \big( J_{n-p}- J_{-n-p}-J_{n+p}+ J_{-n+p}\big)\,\cos(n \tau)
              \right],  
\end{align}
where, here  and also in the  following, all the Bessel  functions are
calculated at the $\Omega_x$ value.The primitive functions become 
\begin{align}
  C_0(\tau) &= J_0 \tau+\sum_{n \geq 1} \big(J_{2n}/n\big)
              \sin(2 n\tau) , \nonumber \\
  S_0(\tau)&=  2\sum_{n \geq 0} \big(J_{2n+1}/(2n+1)\big)(1-\cos ((2n+1)\tau), \nonumber \\
  2 C_1(\tau) &= \cos\Phi (J_{-p} + J_p) \tau + \nonumber \\
            &\sum_{n\geq 1}\left[ \sin \Phi\,\frac{J_{n-p}- J_{-n-p}-J_{n+p}+ J_{-n+p}}{n}\,
              (\cos(n\tau)-1)) \right. \nonumber \\
            &\left. + \cos\Phi\,
              \frac{J_{n-p}+ J_{-n-p}+J_{n+p}+ J_{-n+p}}{n}\,\sin(n \tau)\right],
              \nonumber\\           
  2 S_1(\tau)&= \sin\Phi (J_{-p} - J_p) \tau +\nonumber \\
            & \sum_{n\geq 1}\left[ -\cos \Phi\,
              \frac{J_{n-p}- J_{-n-p}+J_{n+p}- J_{-n+p}}{n}\,
              (\cos(n\tau)-1)) \right. \nonumber \\
            &\left. + \sin\Phi \,
              \frac{J_{n-p}- J_{-n-p}-J_{n+p}+ J_{-n+p}}{n}\,\sin(n \tau)\right].  
              \label{eq:C0S0C1S1}
\end{align}

\section{FMR effective fields}
\label{sec:FMReffectiveFields}
To handle the $\Omega_x$ dressing through the FMR gauge transformation,
we examine here the action of $\Omega_ys_y(t)$ and $\vec{\omega}_0$ on
the spin  evolution applying  the Magnus perturbation  approach. Using
the  functions defined by Eqs.~\eqref{eq:def:csp:coeffs}    we   write   the
$ \vec{h}^{FMR}$ of Eq.~\eqref{eq:def:m} as
\begin{eqnarray}
  \label{eq:m:shorthand}
      \vec{h}^{FMR}(\tau) &=&{}
    \Omega_y
    \begin{pmatrix}
      0\\
      \phantom{-}c_1\\
      -s_1
    \end{pmatrix}
  + 
  \begin{pmatrix}
      1 & 0 & 0 \\
      0 & \phantom{-}c_0 & s_0 \\
      0 & -s_0 & c_0 
    \end{pmatrix}
  \begin{pmatrix}
    \omega_{0x}\\
    \omega_{0y}\\
    \omega_{0z}
  \end{pmatrix} \nonumber \\
  &=& \vec{h}^{FMR}_0(\tau) + \ten{g}_0(\tau) \vec{\omega}_0,  
 \end{eqnarray}
with the  $3\times 3$ $\ten{g}_0(\tau)$ matrix  determining the static
field contribution in analogy to the effective field perturbation expansion.\\ 
\indent We  define the  $\vec{H}^{FMR}$ vector as definite  integral of
$\vec{h}^{FMR}$ 
\begin{equation}
  \label{eq:def:M}
  \vec{H}^{FMR}(\tau) = \int_0 ^{\tau} \vec{h}^{FMR}(\tau') \de \tau'.
\end{equation} 
\indent  For the $\vec{h}^{FMR}(\tau)$    and
$\vec{H}^{FMR}(\tau)$  quantities, also the    mean    values   over  the $(0,2\pi)$ interval are  required
and   denoted   in   the   following   as   $\mv{\vec{h}^{FMR}}$   and
$\mv{\vec{H}^{FMR}}$, respectively.  Notice 
\begin{equation}
 \vec{H}^{FMR}(2\pi) = 2 \pi \mv{\vec{h}^{FMR}},
 \label{eq:expression1}
\end{equation}
and the following connection between mean values:
\begin{equation}
\mv{\tau \vec{h}^{FMR}} = 2\pi \mv{\vec{h}^{FMR}} - \mv{\vec{H}^{FMR}}.
\label{eq:expression2}
\end{equation}
\indent  Similarly for  $\vec{H}^{FMR}(\tau)$, using  the functions
of Eqs.~\eqref{eq:def:CSG:coeffs} we write
\begin{eqnarray}
  \label{eq:M:shorthand}
      \vec{H}^{FMR}(\tau) &=&{}
    \Omega_y
    \begin{pmatrix}
      0\\
      \phantom{-}C_1\\
      -S_1
    \end{pmatrix} + 
     \begin{pmatrix}
      \tau & 0 & 0 \\
      0 & \phantom{-}C_0 & S_0 \\
      0 & -S_0 & C_0 
    \end{pmatrix}
\, 
  \begin{pmatrix}
    \omega_{0x}\\
    \omega_{0y}\\
    \omega_{0z}
  \end{pmatrix}  \nonumber \\
  &=&{}  \vec{H}^{FMR}_0(\tau) + \ten{G}_0(\tau) \vec{\omega}_0 ,
\end{eqnarray}
again with the $\ten{G}_0(\tau)$ $3 \times 3$ matrix characterising the static field dependence.\\
\indent Replacing  the above FMR quantities  within the Fourier-Magnus
expansions of Eqs.~\eqref{eq:F1:L1}, we find
\begin{align}
\Lambda_1 & = \frac{1}{2}\mv{\vec{h}^{FMR}} \cdot \vec{\sigma}, \\
    {\cal K}_1(\tau) &=\left[\frac{1}{2}\int_0^\tau \vec{h}^{FMR}(\tau')\de \tau'       - \tau  \frac{1}{2}\mv{\vec{h}^{FMR}}\right]\cdot \vec{\sigma}   \nonumber \\
    &= \frac{1}{2}\left[ \vec{H}^{FMR}(\tau) - \tau  \mv{\vec{h}^{FMR}}\right]
    \cdot \vec{\sigma}.
\label{eq:Lambda1:form}
\end{align}
Replacing  these  quantities  within  the  $\Lambda_2$  expression  of
Eqs.~\eqref{eq:F1:L1} and using the Eqs.~\eqref{eq:expression1} and ~\eqref{eq:expression2},
 the second order correction results
\begin{equation}
  \begin{split}
    \Lambda_2 
    &= \frac{1}{4}\mv{\vec{h}^{FMR} \times \vec{H}^{FMR}}\cdot\vec{\sigma},    
\end{split}
  \end{equation}
  \indent From  the above $\Lambda_1$ and  $\Lambda_2$ expressions the
  first and  second order effective fields  of Eq.~\eqref{eq:h:pertur}
  result
\begin{equation}
  \label{eq:def:1e2ord}    
    \vec{h}  \approx  \vec{h}^{(1)} + \vec{h}^{(2)} =  \mv{\vec{h}^{FMR}} + \frac{1}{2}\mv{\vec{h}^{FMR} \times  \vec{H}^{FMR}}.
\end{equation}
Using      the     expressions      of     Eqs.~\eqref{eq:m:shorthand}
and~\eqref{eq:M:shorthand}  and performing  some  algebra, the  second
order contribution is rewritten as
\begin{eqnarray}
  \label{eq:mxM}
    \vec{h}^{(2)}  &=&\frac{1}{2} \mv{\vec{h}^{FMR}_0\times\vec{H}^{FMR}_0}  \nonumber  \\ 
    &\phantom{=}& + \frac{1}{2} \left(
      \mv{ \vec{h}^{FMR}_0\times \left[ \ten{G}_0  \vec{\omega}_0 \right]}
      - \mv{\vec{H}^{FMR}_0\times  \left[ \ten{g}_0 \vec{\omega}_0 \right] }
    \right) \nonumber \\
    &\phantom{=}& +\frac{1}{2}
    \mv{\left[ \ten{g}_0 \vec{\omega}_0 \right]
      \times \left[ \ten{G}_0 \vec{\omega}_0 \right] }
\end{eqnarray}
where the three lines give the  contributions of Eqs,~\eqref{eq:h20:synte},
  \eqref{eq:g2ord}  and 
\eqref{eq:Y1}, respctively.

 \section{Cosine second order perturbation}
\label{sec:secondorder}
The  $Q_x$ amplitude of the synthetic field $\vec{h}^{(2)}_{s,x}$ is derived from Eq.~\eqref{eq:qx} inserting the cosine driving $\beta_n$ expressions of Eqs.~\eqref{eq:alphanbetan:prl}. It results
\begin{equation}
  \begin{split}
    Q_{x} &= -\frac{J_p}{2} 
    (1 + (-1)^p \cos(2\Phi))
    \sum_{n \geq 1}\frac{J_{n+p} - J_{p-n}}{n} \\
    &-\frac{J_p}{2}( (-1)^p + \cos(2\Phi))
    \sum_{n \geq 1}\frac{J_{n-p} - J_{-p-n}}{n}, 
    \end{split}
    \label{eq:Qxx:prl} 
\end{equation}
\indent For the $\ten{g}^{(2)}$ tensor components of Eq.~\eqref{eq:g2ord}, inserting the $\alpha_n$ and $\beta_n$ expressions quoted above one finds
\begin{eqnarray}
  Q_{xy} &=& (A + B)\cos(\Phi), \nonumber \\
  Q_{xz} &=& -(A - B)\sin(\Phi), \nonumber \\
  Q_{yx} &=& C \cos(\Phi), \nonumber \\
  Q_{zx} &=& -D \sin(\Phi),
  \label{eq:Qij:prl} 
\end{eqnarray}
where
\begin{eqnarray}
 A & =& \sum_{n \neq 0}\frac{1}{n}
      ( J_{n-p}J_n - J_n J_{-p} -J_0J_{n-p}), \nonumber \\
  B &=&  \sum_{n \neq 0}\frac{1}{n}
      ( J_{n+p}J_n - J_n J_{p} -J_0J_{n+p}), \nonumber \\
  C &=& \sum_{n \geq 1} \frac{1}{n}
      ( J_{p+n} -J_{p-n} + J_{n-p}-J_{-n-p}), \nonumber \\
  D &=& \sum_{n \geq 1} \frac{1}{n}
       ( J_{-p+n} -J_{-p-n} - J_{n+p}+J_{p-n}).
        \label{eq:sommatorie} 
\end{eqnarray}
Finally from $(\alpha_n,\beta_n)$ Bessel function dependencies we obtain for the $\ten{f}^{(2)}$ components of Eqs.~\eqref{eq:def:qc:qs}  
\begin{eqnarray}
q_c &=& 0, \nonumber \\
  q_s &=& \sum_{n=0}^{+\infty}\frac{ J_{2n+1}}{2n+1}, \nonumber \\
  q_0 &=& - J_0\, q_s.
\label{eq:qi:prl}
\end{eqnarray}

\section{Kick operator }
\label{kickoperator}
The kick operator ${\cal K}_1$ given in Eq.~\eqref{eq:Lambda1:form} at
the    first    perturbation     order    is    rewritten    inserting
Eq.~\eqref{eq:M:shorthand} for the $\vec{H}^{FMR}$ vector and deriving
$\mv{\vec{h}^{FMR}}$ from Eq.~\eqref{eq:m:shorthand}. It leads to
 \begin{equation}
  \label{eq:K1:cosine}
      {\cal K}_1(\tau) =
    \frac{\Omega_y}{2}
    \begin{pmatrix}
      0\\
      \tilde{C}_1\\
      -\tilde{S}_1
    \end{pmatrix}\cdot \vec{\sigma}  
  +  \frac{1}{2}
     \begin{pmatrix}
      0 & 0 & 0 \\
      0 & \tilde{C}_0 & \tilde{S}_0 \\
      0 & -\tilde{S}_0& \tilde{C}_0 
    \end{pmatrix}
  \begin{pmatrix}
    \omega_{0x}\\
    \omega_{0y}\\
    \omega_{0z}
  \end{pmatrix}  \cdot  \vec{\sigma},
\end{equation}
where    for   the    cosine    driving    the   following    combined
$(\tilde C_i, \tilde S_i)$ functions are introduced:
\begin{align}
 \tilde{C}_0(\tau) &= \sum_{n \geq 1} \big(J_{2n}/n\big) \sin(2 n\tau), \nonumber \\
 \tilde{S}_0(\tau) &= 2\sum_{n \geq 0} \big(J_{2n+1}/(2n+1)\big)(1-\cos ((2n+1)\tau),\nonumber \\
 \tilde{C}_1(\tau) &= C_1-\tau \mv{c_1} = C_1 - \tau\mathrm{Re}(\beta_0),\nonumber\\  
  \tilde{S}_1(\tau)&= S_1-\tau \mv{s_1} = S_1 -  \tau \mathrm{Im}(\beta_0),
\end{align}
and 
\begin{equation}
\beta_0=\cos(\Phi)(J_p+J_{-p})-i\sin(\Phi)(J_p-J_{-p}).
\end{equation}
\indent For  the class  of single cosine  dressing experiments  as the
original   one   by   Landr\'e  et   al.~\cite{LandreCohen_70}   where
$\Omega_y=0$ and $\omega_{0x}=\omega_{0z}=0$,  the above kick operator
reduces to 
 \begin{equation}
  \label{eq:K1:Landre}
      {\cal K}_1(\tau) 
  =\frac{1}{2}\omega_{0y}\left(\tilde{C}_0\sigma_y-\tilde{S}_0\sigma_z\right).
\end{equation}
\indent    For    the    Cs   dual    cosine    dressing    experiment
of~\cite{Bevilacqua_PRL_20}  where  $\omega_{0y}=0$,  the  above  kick
operator reduces to 
 \begin{equation}
  \label{eq:K1:PRL20}
      {\cal K}_1(\tau) =\frac{\Omega_y}{2}\left(\tilde{C}_1\sigma_y-\tilde{S}_1\sigma_z\right)+
      \frac{\omega_{0z}}{2}\left(\tilde{S}_0\sigma_y+\tilde{C}_0\sigma_z\right).
\end{equation}
For  all   these  experiments   operating  with  low   $\Omega_y$  and
$\vec{\omega}_{0}$     values,     the     spin     micromotion     of
Eq.~\eqref{eq:Mmicrom}  is dominated  by  the first  term  due to  the
transformation to the FMR frame.

\bibliography{biblioDualDressing2}
   
\end{document}